\documentclass[useAMS, usenatbib]{mnras}

\usepackage{amsmath}
\usepackage{graphicx}
\usepackage{rotating}
\usepackage{journal}
\usepackage{multicol}
\usepackage{array}
\usepackage{url}
\newcommand{\sniip}{ASASSN-14dq}
\newcommand{\hostglx}{UGC 11860}
\newcommand{\maghundred}{$\rm mag\ (100 d)^{-1}$}
\newcommand{\magday}{$\rm mag\ d^{-1}$}
\newcommand{\ebv}{$E(B-V)$}
\newcommand{\kms}{$\rm km\ s^{-1}$}

\title[Type II-P SN: \sniip]{\sniip: A fast-declining type II-P Supernova in a low-luminosity host galaxy}
\author[Avinash et al.]{Avinash Singh$^{1,2}$\thanks{e-mail: avinash21292@gmail.com}, S. Srivastav$^3$, Brajesh Kumar$^1$, G.C. Anupama$^1$, D.K. Sahu$^1$\\ \\
$^1$Indian Institute of Astrophysics, Koramangala 2nd Block, Bengaluru - 560034, India\\
$^2$Joint Astronomy Programme, Department of Physics, Indian Institute of Science, Bengaluru - 560012, India\\
$^3$Department of Physics, Indian Institute of Technology, Powai, Mumbai - 400076, India\\}
\begin{document}

\label{firstpage}
\maketitle

\begin{abstract}

Optical broadband (\textit{UBVRI}) photometric and low-resolution spectroscopic observations of the type II-P supernova (SN) \sniip\ are presented. \sniip\ exploded in a low-luminosity/metallicity host galaxy \hostglx, the signatures of which are present as weak iron lines in the photospheric phase spectra. The SN has a plateau duration of $\sim\,$90 d, with a plateau decline rate of 1.38 \maghundred\ in \textit{V}-band which is higher than most type II-P SNe. \sniip\ is a luminous type II-P SN with a peak $V$-band absolute magnitude of --17.7$\,\pm\,$0.2 mag. The light curve of \sniip\ indicates it to be a fast-declining type II-P SN, making it a transitional event between the type II-P and II-L SNe. The empirical relation between the steepness parameter and $\rm ^{56}Ni$ mass for type II SNe was rebuilt with the help of well-sampled light curves from the literature. A $\rm ^{56}Ni$ mass of $\sim\,$0.029 M$_{\odot}$ was estimated for \sniip\, which is slightly lower than the expected $\rm ^{56}Ni$ mass for a luminous type II-P SN. Using analytical light curve modelling, a progenitor radius of $\rm \sim3.6\times10^{13}$ cm, an ejecta mass of $\rm \sim10\ M_{\odot}$ and a total energy of $\rm \sim\,1.8\times 10^{51}$ ergs was estimated for this event. The photospheric velocity evolution of \sniip\ resembles a type II-P SN, but the Balmer features (H$\alpha$ and H$\beta$) show relatively slow velocity evolution. The high-velocity H$\alpha$ feature in the plateau phase, the asymmetric H$\alpha$ emission line profile in the nebular phase and the inferred outburst parameters indicate an interaction of the SN ejecta with the circumstellar material (CSM).

\end{abstract}


%------------------------------------------------------------------------------%
\section{Introduction}
\label{sec:intro}
%------------------------------------------------------------------------------%

Core collapse supernovae (CCSNe) are catastrophic events that mark the end of stellar evolution in massive stars \citep[M $\geq\,$8 $\rm M_\odot$,][]{2003heger}. CCSNe result when the thermal energy content of the star is unable to prevent self-gravitational collapse, i.e at the end of nuclear burning. CCSNe are broadly divided into two types: hydrogen-rich (II-P, II-L, IIb, IIn) and hydrogen-deficient (Ib, Ic, Ibn) \citep{1941minkowski, 1997filippenko}. The most common sub-type of CCSNe is II-P \citep{2011li}, which results from a star that retained most of its hydrogen envelope at the time of the explosion resulting in the presence of prominent hydrogen Balmer lines in the spectra. Light curves of these SNe are characterized by a distinctive flat spread (called the \lq plateau\rq phase) where the luminosity stays almost constant for an extended period of time \citep[$\sim\,$100 days]{1979barbon}. During the shock breakout, the hydrogen in the envelope gets ionized. As the envelope expands, the recombination wave moves inwards (in mass), staying at roughly the same radius and temperature. Recombination provides the major source of energy by allowing the stored energy to be radiated \citep{2010dessart} once the envelope cools enough for the hydrogen to recombine. The recombination process competes with the homologously expanding (and hence cooling) ejecta, resulting in the presence of a plateau phase in a type II-P SN. We use the definition of plateau duration as the optically thick phase duration (OPTd), which spans from the time of the explosion till the start of the transition to the radioactive tail \citep{2014anderson}.

At the end of the recombination phase i.e. when the recombination front meets the inner denser material, the SN settles onto a slowly declining radioactive phase after experiencing a significant drop in luminosity, which depends on the amount of radioactive $\rm ^{56}Ni$ synthesized in the explosion. Type II SNe generally follow the decay rate of $\rm ^{56}Co$ in the nebular phase \citep{1990turatto}. Based on the amount of hydrogen in the envelope, type II-P SNe show a wide variety in plateau duration, plateau luminosity and expansion velocities \citep{2003hamuy, 2014anderson, 2014afaran}. These observational parameters are directly connected to the explosion parameters and can help infer the properties of the progenitor star \citep{2015sanders}.

Hydrodynamical modelling of some well-studied type II-P SN light curves (LC) have suggested that a red supergiant (RSG) progenitor with an extended hydrogen envelope is essential for producing the plateau in the light curves \citep{1971grassberg, 1976chevalier}. Modelling of such SNe has predicted progenitors with masses ranging from 15-25 $\rm M_{\odot}$ \citep{2011bersten, 2015morozova} although, the estimates of masses of progenitors from direct imaging has been limited to 9--17 $\rm M_{\odot}$ \citep{2009smartt}.

Type II-P SNe have a great potential as a standard candle (next only to type Ia) for extra-galactic distance estimation owing to their well-studied theoretical models along with the extensive observational study performed on a large sample of them. The two most commonly used techniques are the standard candle method \citep[SCM]{2002hamuy} and the expanding photosphere method \citep[EPM]{1974kirshner}, which was improved later by \citep{2001hamuy, 2005adessart}. The SCM is based on an observationally determined relation between luminosity and the expansion velocity of a type II-P SN. The EPM relates the photospheric and the angular radii of a supernova in order to derive its distance. The effect of progenitor metallicity on type II SNe spectra for a standard Red Supergiant (RSG) explosion has been studied in the past \citep{2005bdessart, 2009kasen, 2013bdessart}. Metal lines appearing in type II SNe during the recombination phase helps constrain metallicity of the progenitor star as the region probed during its photospheric evolution retains the composition of the progenitor \citep[][hereafter D13, D14 and A16, respectively]{2013bdessart, 2014dessart, 2016anderson}. Type II SNe arising from progenitors with lower metallicity display fewer and weaker (low Equivalent-Width, EW) metal lines.

Although type II-P SNe were originally considered to be part of a different population than the type II-L, the distribution of type II SNe seems to form a continuous distribution according to many authors \citep{1994patat, 2014anderson, 2015sanders}. \citet{2012arcavi} pointed out the presence of type II-L and type II-P SNe as two contrasting populations based on their $R$-band light curve. \citet{2014afaran, 2014bfaran} also pointed towards the differences between the two sub-types through their spectroscopic evolution. \citet{2014bfaran} highlighted the slow evolving H$\beta$ line velocities and the presence of weak hydrogen absorption lines in type II-L SNe in comparison with type II-P SNe, although they agreed with the continuum of photometric features seen in type II SNe. The primary characterizing difference between the two sub-types is the fast declining light curve in a type II-L SN, which is seen as a result of less amount of hydrogen ($\sim\,$1-2 $\rm M_{\odot}$) in the envelope \citep{1993blinnikov}. The inference of \citet{2014bfaran} also signify a hydrogen-poor envelope in type II-L SNe and is consistent with the excessive mass-loss rate associated with the higher Zero-Age main sequence (ZAMS) mass of a type II-L SN progenitor compared to that of a type II-P progenitor \citep{2010elias, 2011elias}. However, \citet{2015maund} discovered that the previously thought progenitor of type II-L SN 2009kr \citep{2010elias} is actually a small compact cluster.

\citet{2017morozova}, based on numerically investigating the broadband light curves of RSGs with CSM argued that type II-L SNe might also result from the presence of dense CSM close to the progenitor star. They also highlight the fact that the sharp density gradient between the wind and the underlying RSG model is required for high decline rate in the light curves, and that the large radius of the progenitor is not sufficient enough to reproduce that change. The presence of continuity in the wind properties of progenitor stars show agreement with the continuum of light curve properties seen in type II SNe \citep{2014anderson}.

In order to classify type II SNe, \citet{2014bfaran}, defined a parameter s$\rm 50_V$, which is the magnitude decline in the $V$-band light curve between the maximum and 50 days after the explosion. Type II-P SNe were classified as having an s$\rm 50_V\leq$ 0.5. \citet{2015valenti} also differentiated the two sub-types using the same parameter but with a higher cut-off, i.e  s$\rm 50_V\,\leq\,$1.0 for type II-P SNe. \citet{1994patat} differentiated the sub-types based on their $B$-band decline rates and classified type II SNe with $\rm \beta^{B}_{100}\,<\,3.5$ $\rm mag (100\ d)^{-1}$ as type II-P SNe. Type II-L SNe are also, in general, more luminous (at maximum) compared to type II-P SNe on an average by $\sim\,$1.5 mag \citep{1994patat, 2002richardson, 2014anderson}. Using the data from Lick Observatory Supernova Search (LOSS), \citet{2011li} inferred an average absolute magnitude (at maximum) of --15.66$\,\pm\,$0.16 ($\sigma$\,=\,1.23) for type II-P SNe and --17.44$\,\pm\,$0.22 for type II-L SNe. \citet{2014anderson} also computed an average peak $V$-band magnitude of --16.74 mag ($\sigma$\,=\,1.01 mag) for type II SNe.

Due to an increased number of sky surveys, the detection of SNe (and hence type II SNe) in the recent past have increased. With the increased amount of published light curves showing a continuum of properties, the classification of type II SNe still remains an open question. \citet{2015valenti} suggest that a clear distinction in the masses of the hydrogen envelope of type II-P and type II-L SNe will direct towards the possibility of a phase in stellar evolution that strips a star of discrete chunks of hydrogen. This motivates the observational monitoring and detailed analysis of more type II SNe in order to better understand the distinction between the two sub-types. \sniip\ is interesting in the fact that it exploded in a low-luminosity host galaxy \hostglx\, unlike most type II SNe.

\begin{figure}
\centering
\resizebox{\hsize}{!}{\includegraphics[width=\linewidth]{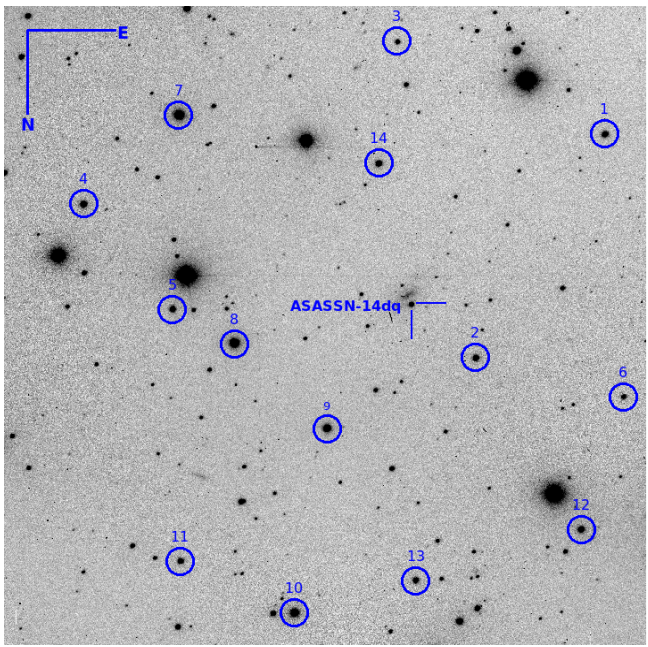}}
\caption{\sniip\ in \hostglx. The $V$-band image shown above covers a subsection about 9\arcmin $\times$ 9\arcmin\ and is taken from 2m HCT. The circled stars (numbered 1-14) in the field refer to the secondary standard stars used for calibration (see Table~\ref{tab:secstd} for their magnitudes).}
\label{fig:findingchart}
\end{figure}

\sniip\ was discovered in images obtained on 2014 July 08.48 UT in the dwarf galaxy \hostglx\ by the quadruple 14-cm \lq \lq Brutus\rq \rq\ telescope in Haleakala, Hawaii of the All Sky Automated Survey for SuperNovae (ASAS-SN) and was reported in the \textit{Astronomer's Telegram} \citep{2014discovery}. \sniip\ lies $\sim\,$9\arcsec($\sim\,$1.6 kpc) from the center of the host galaxy \hostglx\ (z\,=\,0.010, luminosity distance d\,=\,38.5 Mpc). The transient was reported to have an absolute $V$-band magnitude of approximately $-17.0$ on discovery. Spectroscopic data obtained \citep{2014classify} on 2014 July 9.5 UT with the FLOYDS low-resolution spectrograph mounted on the Faulkner Telescope North displayed a blue continuum having broad hydrogen P-Cygni features with an absorption minimum at $\sim\,$10000 \kms. Superfit \citep{2005howell}, a supernovae spectrum fitting code matched the above spectrum with an early spectrum of the type IIb SN 1993J at the redshift of the host galaxy, but it was too early to be sure of the precise sub-type of \sniip.

In this paper, we present detailed photometric and spectroscopic analysis of \sniip\ having a very well-sampled photometric coverage during the transitional phase, which is rarely the case because of its short duration ($\sim\,$10 d). In Section~\ref{sec:obs}, we describe the optical observations of \sniip\ along with a brief outline of the data reduction procedure. In Section~\ref{sec:hostglx}, we discuss the properties of the host galaxy, and estimate the reddening along the line-of-sight and the distance to the host galaxy. Spectroscopic analysis of \sniip\ is presented in Section~\ref{sec:specevol}. We rebuild the empirical relation between the steepness parameter and $\rm ^{56}Ni$ mass for type II SNe and estimate $\rm ^{56}Ni$ mass for \sniip\ in Section~\ref{sec:nickelmass}. We fit the bolometric light curve of \sniip\ using an analytical model and extract outburst parameters in Section~\ref{sec:lcmodel}. We compare \sniip\ with well-studied type II SNe from literature in Section~\ref{sec:disc}. We summarize the results obtained in this paper in Section~\ref{sec:sum}.

\begin{table}
\label{tab:hostgal}
\caption{Parameters of \sniip\ and its host galaxy.}
\renewcommand{\arraystretch}{1.1}
\setlength{\tabcolsep}{2.5pt}
\begin{tabular}{l l c}
\hline \hline 
\noalign{\smallskip}
Parameters 			    & Value		 				                        & Ref.\\

\noalign{\smallskip} \hline \noalign{\smallskip} 
\multicolumn{3}{l}{\textit{\underline \sniip}:} \\ \\

RA (J2000) 			    & $\alpha\,=\,21^{\rm h} 57^{\rm m} 59\fs97$ 	    & 3 \\
DEC (J2000)             & $\delta\,=\,+24\degr 16\arcmin 08\farcs1$ 	    & 3 \\
Galactocentric Location	& 3\farcs7 E, 8\farcs1 N 			                & 3 \\ \\

Discovery Date 		    & $t_{\rm d}$\,=\,2014 July 08.48 (UTC) 		    & 3 \\
			            & (JD 2456846.98)            			            &   \\
Explosion Date 		    & $t_{\rm 0}$\,=\,2014 July 03.00 (UTC) 		    & 1 \\
			            & (JD 2456841.50$\,\pm\,$5.5)    			        &   \\ \\
			            
Peak Magnitude 		    & $M_{V}$\,=\,--17.7$\,\pm\,$0.2 mag                & 1 \\
Total reddening         & $E(B-V)$\,=\,0.0601$\,\pm\,$0.0006 mag            & 1 \\

\noalign{\smallskip} \hline \noalign{\smallskip}
\multicolumn{3}{l}{\textit{\underline \hostglx}:} \\ \\

Alternate name 		    & MCG +04-51-014, PGC067733                         & 2 \\
Type                    & Sdm 						                        & 4 \\
RA (J2000) 			    & $\alpha\,=\,21^{\rm h} 57^{\rm m} 59\fs82$ 	    & 2 \\
DEC (J2000)             & $\delta\,=\,+24\degr 15\arcmin 59\farcs7$ 	    & 2 \\ \\

Red-shift               & z\,=\,0.010424$\,\pm\,$0.000010                   & 2 \\
Luminosity Class        & LC IV-V                                           & 4 \\
Luminosity Distance     & D\,=\,44.8$\,\pm\,$3.1 Mpc                        & 1 \\
Distance modulus        & $\mu$\,=\,33.25$\,\pm\,$0.15 mag 			        & 1 \\
\noalign{\smallskip} \hline
\end{tabular}
\newline 
(1) This paper;
(2) \citet{1996hostgal};
(3) \citet{2014discovery};
(4) \citet{1991devac}.
\end{table}

%------------------------------------------------------------------------------%
\section{Observation and Data Reduction}
\label{sec:obs}
%------------------------------------------------------------------------------%

\subsection{Photometry}

Photometric monitoring of \sniip\ started on 2014 July 15 (JD 2456854.3). The broadband photometric observations in Bessell $UBVRI$ bands were carried out at 34 epochs using the Himalayan Faint Object Spectrograph Camera (HFOSC), mounted on the 2-m Himalayan Chandra Telescope (HCT), situated at the Indian Astronomical Observatory (IAO), Hanle, India. The camera is equipped with a SITe CCD chip having a dimension of 2048 $\times$ 4096 pixels. The readout noise and gain of the camera are 4.87$e^{-}$ and 1.22 $e^{-}$/ADU, respectively. The central 2048 $\times$ 2048 pixels of the CCD cover a field-of-view (FOV) of 10\arcmin $\times$ 10\arcmin\ and was used for photometric observations. Several bias and sky flat frames were obtained in addition to the object frames. Data reduction software IRAF\footnote{IRAF is distributed by the National Optical Astronomy Observatory, which is operated by the Association of Universities for Research in Astronomy (AURA) under a cooperative agreement with the National Science Foundation.} was used for performing routine pre-processing tasks of bias-subtraction, flat-fielding, and cosmic ray removal. During the nebular phase of the supernova, multiple object frames were obtained in the same band and were combined (after aligning) to improve the signal-to-noise ratio (SNR) in the resultant object frame.

The secondary standard stars in the SN field were calibrated using Landolt photometric standards \citep{1992landolt} PG0231+051, PG0918+029, and PG0942-029. The Landolt standards and the SN field were observed on two photometric nights 2014 November 17 and 23, to obtain transformation coefficients. Aperture photometry was performed on the stars in the standard fields and the SN field using the \textit{phot} task in IRAF. The photometry was done at two different apertures - FWHM of the stellar profile and 4 times the FWHM of the stellar profile, to calculate the aperture correction.

The average atmospheric extinction coefficients for the site, in each of the Bessell $U$, $B$, $V$, $R$ and $I$ bands were adopted from \citet{2008stalin}. The zero points were determined using the average colour terms for the telescope detector system. These were used to calibrate the secondary standards in the SN field. The calibrated $UBVRI$ magnitudes along with the  $1\sigma$ uncertainty of the secondary standards in the SN field are mentioned in Table~\ref{tab:secstd}. The magnitudes of \sniip\ were determined using point-spread function (PSF) photometry available with the DAOPHOT package in IRAF.

The last non-detection of \sniip\ was in images taken on 2014 June 27.49 UT. Therefore, we adopt the date mid-way between the non-detection and detection of the supernova as its explosion date, JD\,=\,2456841.50 $\,\pm\,$ 5.5 (i.e. 2014 July 03.0 UT).

\subsection{Spectroscopy}

The spectroscopic monitoring of \sniip\ from HCT started on 2014 July 16 (JD 2456855.4). Low-resolution spectroscopic observations were carried out on 23 epochs with the HFOSC using grisms Gr7 (3500--7800 \AA) and Gr8 (5200--9250 \AA). Bias subtraction and cosmic ray correction were performed on all the raw two-dimensional images. One-dimensional spectra were extracted optimally \citep{1986horne} from the original two-dimensional images using the package \textit{twodspec} in IRAF. Arc lamp spectra FeAr and FeNe were used for wavelength calibration. Night sky emission lines $\lambda$5577, $\lambda$6300 and $\lambda$6363 were used to cross-check the wavelength calibration and small shifts were applied wherever necessary. Instrumental response curves were generated using spectra of spectrophotometric standards (Feige110 and Feige 34) and were used for flux calibrating the supernova spectra. The nights on which the standards were not observed, the response curves from the adjacent nights were used. Flux calibrated spectra were scaled using magnitudes obtained through photometric observations from HCT and brought to an absolute flux scale. For the epochs with observations in both the grisms, a single flux calibrated one-dimensional spectrum was generated by combining the spectra in the two grisms with a common overlap region. The telluric lines were not removed from the spectra.

%-----------------------------------------------------------------------------%
\section{Host-galaxy - \hostglx}
\label{sec:hostglx}
%-----------------------------------------------------------------------------%

\begin{table*}
\centering
\caption{Estimates of metallicity of the host galaxy \hostglx.}
\setlength{\tabcolsep}{8pt}
\label{tab:hostmetal}                     
\begin{tabular}{l c c c c c c}          
\hline \hline
Technique   &  Filter       &   Magnitude               &   Color       &   Color Value &   12 + log(O/H)       & Reference     \\
            &               &   (mag)                   &               &   (mag)       &   (dex)               & (Technique)   \\
\noalign{\smallskip} \hline \noalign{\smallskip}
LZ$_1$      &   $B$         &   -17.50$\,\pm\,$0.52     &   --          &   --          &   8.20$\,\pm\,$0.39   &   2           \\
LZ$_2$      &   $B$         &   -17.50$\,\pm\,$0.52     &   --          &   --          &   8.48$\,\pm\,$0.04   &   1           \\
LCZ         &   $i$         &   -17.34$\,\pm\,$0.15     &   $i-z$       &   0.23        &   8.48$\,\pm\,$0.07   &   3           \\
LCZ         &   $u$         &   -15.74$\,\pm\,$0.15     &   $u-g$       &   1.25        &   8.43$\,\pm\,$0.07   &   3           \\
\noalign{\smallskip} \hline \noalign{\smallskip} 
    Mean    &   --      &    --                     &   --       &               &   8.40$\,\pm\,$0.18   &               \\
\noalign{\smallskip} \hline \noalign{\smallskip} 
\end{tabular}
\newline
(1) \citet{2004tremonti};
(2) \citet{2012berg};
(3) \citet{2013sanders}.
\end{table*}

\sniip\ was discovered in the host galaxy \hostglx, also known as PGC067733 and MCG +04-51-014. The galaxy is located at a redshift of 0.010424, which was inferred using the 21-cm neutral hydrogen line measurements \citep{1998theureau}.

\subsection{Line-of-sight Extinction}
\label{sec:losext}

To derive the explosion parameters of the supernova, reddening along the line-of-sight (LOS) of the \sniip\ due to the Milky Way (MW) and the host galaxy (\hostglx) must be estimated. Galactic reddening was obtained from IRSA\footnote{NASA/IPAC Infrared Space Archive \url{http://irsa.ipac.caltech.edu/applications/DUST/}} using the dust-extinction map by \citet{2011schlafly}, which is a re-calibration of the dust map by \citet{1998schlegel}. The re-calibration assumes \citet{1999fitzpatrick} extinction law and states that for a low value of reddening, the values obtained from \citet{1998schlegel} are generally an overestimate to the Milky Way reddening. A Galactic reddening of \ebv\,=\,0.060$\,\pm\,$0.001 mag is estimated by \citet{2011schlafly} in the direction of \sniip. 

The value of reddening in the direction of the SN was also computed by measuring EW of the interstellar Na\,I D absorption feature \citep{2012poznanski} present in the early phase spectra of \sniip. The EW of the Na\,I D absorption feature was found to be 0.20 \AA\ in the spectrum taken on 2014 July 30. Using the relation given in \citet{1990barbon}, we obtain an \ebv\,=\,0.05 mag, which is in good agreement with the value from the dust map. To estimate the host galaxy reddening, we looked for the narrow Na\,I D interstellar absorption feature at the redshift of the host galaxy in the spectrum with the highest SNR in our data-set. The Na\,I D feature at the redshift of the host galaxy \hostglx\ is barely distinguishable from the continuum and is consistent with the absence of Na\,I D feature in the spectra of \sniip\ in \citet{2016valenti}. Although estimating dust extinction using Na\,I D is unreliable and could also be a result of the low resolution of the spectra \citep{2011poznanski}, undetectable Na\,I D features are typically representative of minimal reddening from the host \citep{2013phillips}. Hence, we infer that the host galaxy reddening is negligible compared to the Galactic reddening.

We used the \lq \lq colour method\rq \rq\ by \citet{2010olivares} to further constrain the reddening \ebv\ due to the host galaxy. The technique assumes that the the intrinsic $(V-I)$ colour at the end of the plateau phase is a constant for type II-P SNe, as they should reach the same hydrogen recombination temperature during that phase. The host colour excess was found from the observed $(V-I)$ colour, which is related to the visual extinction through the relation by \citet{2010olivares}:

\begin{eqnarray}
 \rm A_v(V-I) = 2.518[(V-I) - 0.656], \nonumber \\
 \rm \sigma_{A_v} = 2.518 \sqrt{\sigma^2_{(V-I)} + 0.053^2 + 0.059^2}.
\end{eqnarray}

The mean observed $(V-I)$ colour at the end of the plateau phase ($\sim\,$86.5 d) after correcting for Galactic reddening gives an $A_{V_{host}}$\,=\,0.33$\,\pm\,$0.21 mag. This corresponds to an $E(B-V)_{host}$\,=\,0.11$\,\pm\,$0.07 mag, assuming $R_V$\,=\,3.1, which is in contrast with the absence of the Na\,I D absorption at the redshift of the host galaxy. This discrepancy can be explained from the significant differences in hydrogen recombination temperature occurring across type II-P SNe during the plateau phase depending on their H/He abundance ratio \citep{1996arnett}. Hence, the observed colour excess corrected for Galactic reddening might not necessarily result in a reliable estimate of the host galaxy extinction. As a result, we infer the contribution to the total reddening to be purely Galactic in origin and use an \ebv=0.06 mag for total reddening.

\subsection{Distance}
\label{sec:dist}

Various distance estimates to the host galaxy \hostglx\ are available on the NED\footnote{\url{https://ned.ipac.caltech.edu}} website. Due to the absence of a redshift-independent distance estimate, we use the Hubble flow distances estimated for \hostglx, assuming $\rm H_0$\,=\,70 \kms $\rm Mpc^{-1}$. The distance estimates to \hostglx\ in the literature are: 45.7$\,\pm\,$3.2 Mpc after correcting for Virgo infall only and 44.9$\,\pm\,$3.1 Mpc after correcting for Virgo and Great Attractor in-fall \citep{2000virgo}; 46.8$\,\pm\,$3.3 Mpc after correcting for Local Group velocity \citep{1996hostgal} and 38.2$\,\pm\,$2.7 Mpc from the Cosmic Microwave Background (CMB) dipole model \citep{1996cmb}. 

To estimate the distance using the observed supernova parameters, we use the SCM technique. SCM comes from a correlation of the expansion velocities of the ejecta of type II-P SNe with the bolometric luminosity in the plateau phase \citep{2002hamuy}. For our calculations, we use the refined SCM method proposed by \citet{2006nugent}, where $(V-I)$ colour used to perform extinction correction is measured during the mid-plateau phase and not at the end of the plateau phase. The expected absolute magnitude is obtained from the following relation:

\begin{equation}
 \rm M_{I} = -\alpha\ log_{10}(v_{Fe\,II}/5000) - 1.36[(V-I) - (V-I)_{0}] - M_{I_{0}},
\end{equation}

where $\alpha$\,=\,6.69$\,\pm\,$0.50, $(V-I)_{0}$\,=\,0.53 and $M_{I_{0}}$\,=\,17.49$\,\pm\,$0.08 mag.
Using mid-plateau expansion velocity, $\rm v_{Fe\,II}$\,=\,4910$\,\pm\,$100 \kms, mid-plateau $(V-I)$ colour of 0.63$\,\pm\,$0.02 and mid-plateau apparent magnitude, $m_{I}$\,=\,15.80$\,\pm\,$0.02, we obtain a distance modulus, $\rm \mu$\,=\,33.37$\,\pm\,$0.17 corresponding to a distance of 47.2$\,\pm\,$3.6 Mpc. Henceforth, we use the mean of all available distances, \textit{i.e.} 44.8$\,\pm\,$3.1 Mpc ($\rm \mu$\,=\,33.25$\,\pm\,$0.15 mag) as the distance estimate to the source. All the distance estimates to the host galaxy \hostglx\ are mentioned in Table~\ref{tab:dist_host}.

\subsection{Properties of the host galaxy}
\label{sec:prophost}

\begin{figure}
\centering
\resizebox{\hsize}{!}{\includegraphics{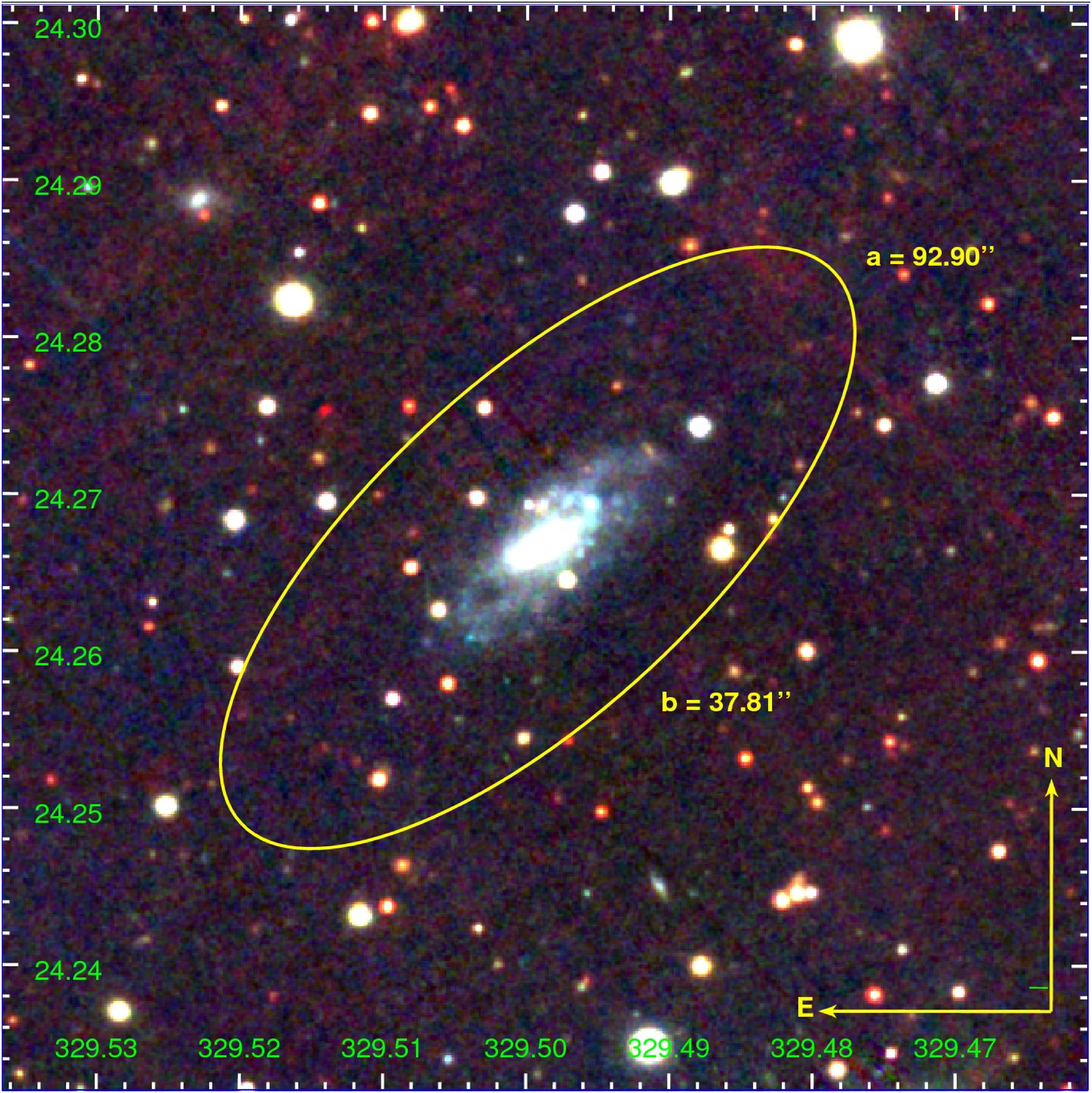}}
\caption{Colour composite image (RGB) of the host galaxy \hostglx\ constructed using images from filters $z$ (red), $r$ (green) and $g$ (blue) obtained from the Pan-Starrs1 Survey (\url{http://ps1images.stsci.edu/cgi-bin/ps1cutouts}). The component images were background subtracted and normalized \textit{w.r.t.} exposure time before combining. The image covers a cropped FOV of $\rm 4\arcmin \times\ 4\arcmin$. The ellipse describes the shape of the galaxy at the isophotal level, $B$\,=\,25 $\rm mag\ arcsec^{-2}$. The labelled values of the axis diameters were taken from NED.}
\label{fig:hostgalrgb}
\end{figure}

\begin{figure}
\centering
\resizebox{\hsize}{!}{\includegraphics{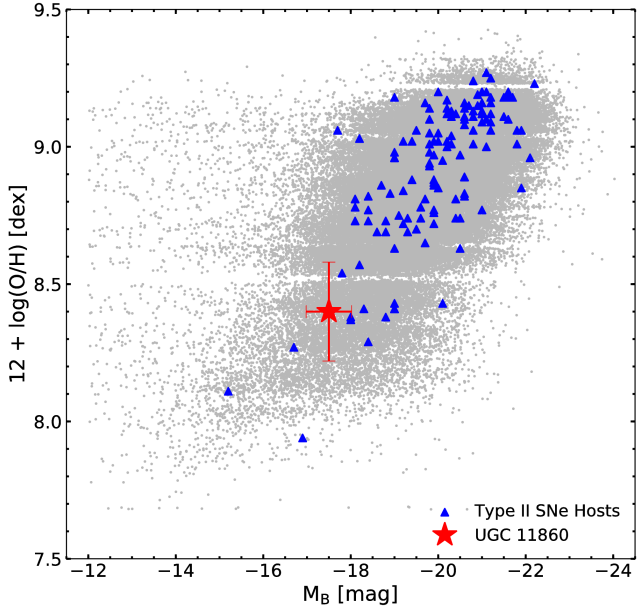}}
\caption{Oxygen abundances of SDSS-DR4 catalog of galaxies ($grey$ dots) used in \citet{2004tremonti} is plotted against their absolute $B$-band magnitude. The $blue$ triangles indicate the hosts of type II SNe from \citet{2008prieto}.}
\label{fig:galmetal}
\end{figure}

\hostglx\ is a characteristically asymmetric and completely bulgeless late-type spiral with an unclear spiral structure, classified as an \lq \lq Sdm\rq \rq\ in the RC3 \citep{1991devac}. The photometric details of the galaxy were obtained from NED and are mentioned in Table~\ref{tab:hostgal}. A mean surface brightness of 24.70 $\rm mag\ arcsec^{-2}$ classifies \hostglx\ as a low-surface-brightness galaxy. \hostglx\ has an absolute $B$-band magnitude of --17.50$\,\pm\,$0.52 and absolute $r$-band magnitude of --17.25$\,\pm\,$0.15, which characterizes it as a low-luminosity dwarf galaxy \citep{2010arcavi}. The galaxy's polar axis is inclined at an angle of $\sim\,$75 degrees with the line-of-sight. Although the nucleus of the galaxy is not well-defined, using the centre as the region of highest surface brightness in the $B$-band image, the SN exploded 3\farcs7 East and 8\farcs1 North from the centre. The isophotal diameter (at $B$\,=\,25 $\rm mag\ arcsec^{-2}$) is 92.9$\farcs$ \citep{1991devac} and corresponds to $\sim\,$20 kpc at the distance of the galaxy, which labels it as a large late-type spiral (see Fig.~\ref{fig:hostgalrgb}).

As no direct metallicity measurements for \hostglx\ are available in the literature, we estimate metallicity of \hostglx\ using various luminosity-metallicity relations. \citet{2012berg} computed oxygen abundances for low-luminosity galaxies ($\rm -10.8\,\geq\,M_B\,\geq\,-18.8$) through a luminosity-metallicity (LZ$_1$) relation using their absolute $B$-band magnitudes (Eqn.~\ref{eqn:lummetal1}). A luminosity-metallicity (LZ$_2$) relation (Eqn.~\ref{eqn:lummetal2}) was also obtained by \citet{2004tremonti} for low-redshift (z$\,\leq\,$0.1) star-forming galaxies using a sample of roughly $\sim\,$53,000 SDSS\footnote{Sloan Digital Sky Survey} galaxies.

\begin{eqnarray}
\label{eqn:lummetal1}
    \rm 12 + log(O/H) = (6.27 \pm 0.21) + (-0.11 \pm 0.01)\times M_B \\
\label{eqn:lummetal2}
    \rm 12 + log(O/H) = (-0.185 \pm 0.001)\times M_B + 5.238 \pm 0.018
\end{eqnarray}

We used the aforementioned relations to compute the oxygen abundance of the host galaxy \hostglx. We also estimated the host oxygen abundance through the luminosity-colour-metallicity (LCZ) relation of \citet{2013sanders} which has less intrinsic scatter compared to the LZ$_2$ relation. This is due to the inclusion of the colour term which correlates highly with galaxy mass-to-light ratio and helps reduce the scatter in the relation occurring due to differences in star formation rate (SFR) of galaxies. \hostglx\ fits the selection criteria used by \citet{2013sanders} for their sample \textit{i.e.} the Galactic extinction corrected $r$-band magnitude of \hostglx\ is less (brighter) than 17.77 mag. We used the magnitudes of \hostglx\ from the SDSS Data Release-14 \citep{2018SDSS} to compute its oxygen abundance. 

The host oxygen abundance obtained through different techniques and their mean value is mentioned in Table~\ref{tab:hostmetal}. The mean oxygen abundance obtained is 8.40$\,\pm\,$0.12 and is lower than the solar value of 8.69$\,\pm\,$0.05 \citep{2009asplund}. Due to the existence of metallicity gradients in galaxies \citep{2004pilyugin}, we expect our metallicity estimate to act as an upper bound for the metallicity estimate of the host environment (and/or the progenitor) as \sniip\ lies at a projected distance of $\sim\,$1.6 kpc away from the centre of the \hostglx. In Fig.~\ref{fig:galmetal}, we plotted the oxygen abundance of galaxies from the SDSS DR4 \citep{2006adelman} catalog against their $B$-band absolute magnitude. In comparison with the host galaxies of type II SNe \citep{2008prieto}, we see that \hostglx\ is among the few hosts with a low oxygen abundance and lies significantly below the majority of the hosts, which mostly have a super-solar oxygen abundance ($\geq\,$8.69). 

%------------------------------------------------------------------------------%
\section{Optical Light Curve}
\label{sec:lightcurve}
%------------------------------------------------------------------------------%

\subsection{Apparent magnitude light curves}
\label{sec:applc}

\begin{table*}
\centering
\renewcommand{\arraystretch}{1.1}
\caption{Light curve parameters of \sniip.}
\label{tab:lcpar}
\begin{tabular}{c c c c c c}
\hline \hline
Filter      &   Plateau slope, s1   &   Plateau slope, s2   &   Avg. Plateau slope  &   Nebular phase slope &   Steepness Parameter \\
            &   (\maghundred)       &   (\maghundred)       &   (\maghundred)       &   (\maghundred)       &   (\magday)           \\
\hline \noalign{\smallskip}            
$U$         &   8.78                &   4.08                &   5.51                &   --                  &   --                  \\       
$B$         &   4.64                &   2.16                &   2.99                &   0.70                &   --                  \\
$V$         &   1.80                &   1.18                &   1.38                &   0.96                &   0.13                \\
$R$         &   0.90                &   0.88                &   0.91                &   0.91                &   --                  \\
$I$         &   0.70                &   0.74                &   0.73                &   1.10                &   --                  \\
\hline
\end{tabular}
\end{table*}

\begin{figure*}
\centering
\resizebox{\hsize}{!}{\includegraphics{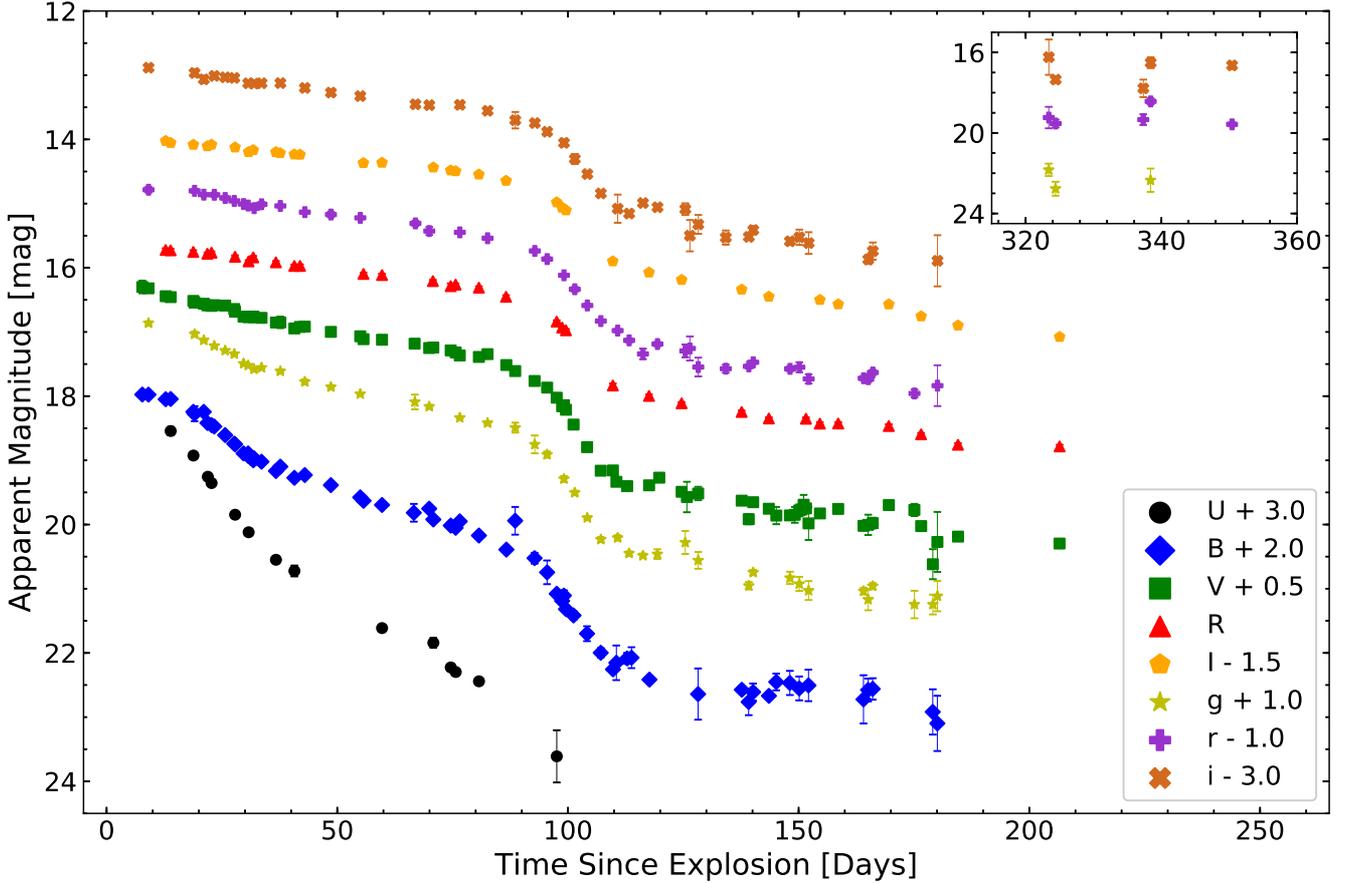}}
\caption{Apparent magnitude light curves of \sniip\ in Bessell $UBVRI$ \& SDSS $gri$ filters. $UBVRI$ data taken from HCT is combined with $BVgri$ data from \citet{2016valenti}. Light curves are plotted with an offset for clarity.}
\label{fig:apparentlc}
\end{figure*}

\begin{figure}
\centering
\resizebox{\hsize}{!}{\includegraphics{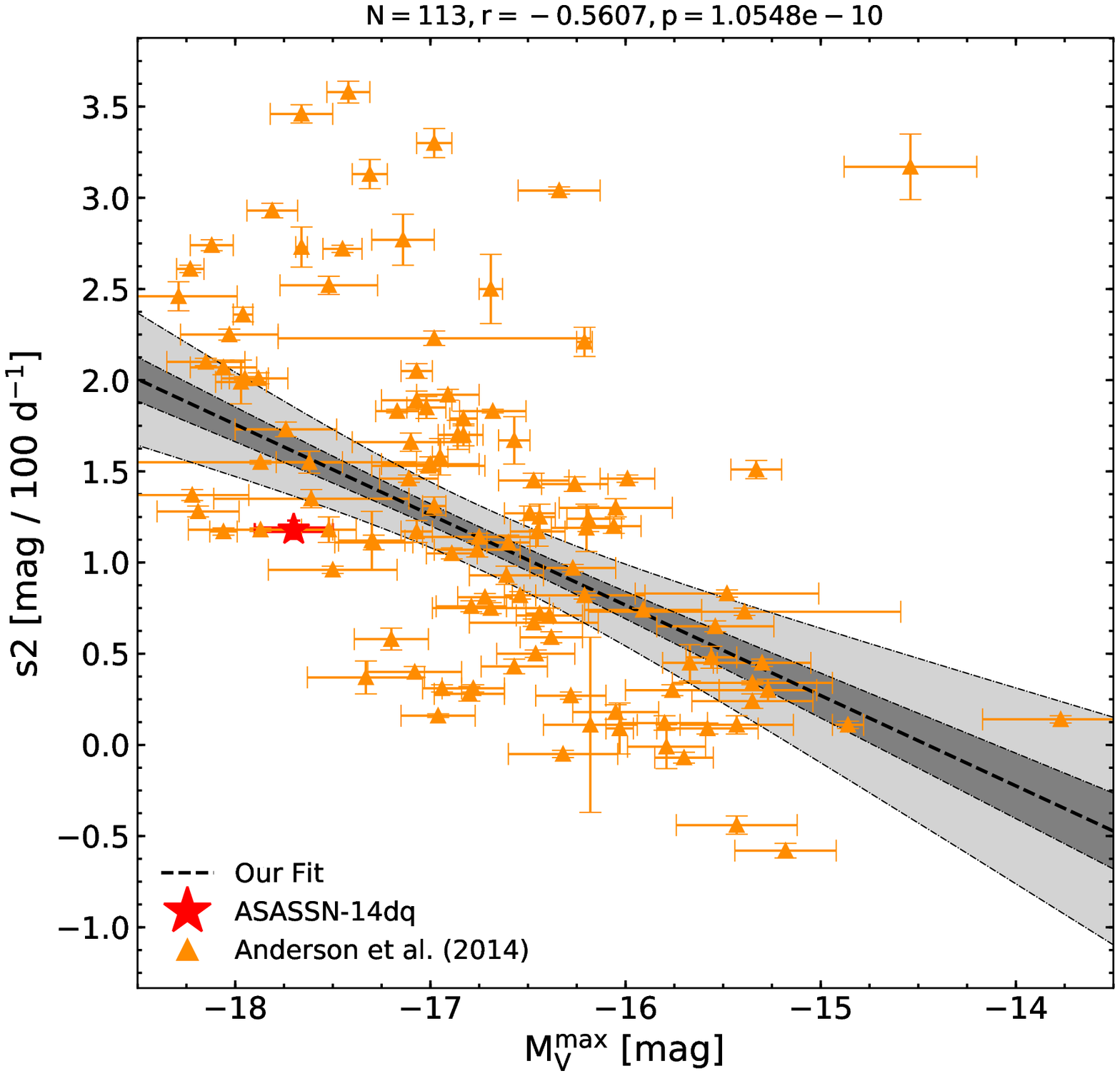}}
\caption{Decline rate during the late-stages of the plateau phase ($s2$) plotted against the peak $V$-band absolute magnitude ($\rm M_V^{max}$). The 1$\sigma$ and 3$\sigma$ confidence intervals of the fit are shaded in $dark-grey$ and $light-grey$, respectively.}
\label{fig:mmaxs2}
\end{figure}

$UBVRI$ broadband photometric observations of \sniip\ were carried out during $+13$ d to $+207$ d since the date of explosion (JD 24568541.50) using the HCT. Our observations were combined with $BVgri$ observations of \sniip\ by the Las Cumbres Optical Global Telescope network \citep[LCOGT,][]{2016valenti}. The apparent magnitude light curves after combining the data from both the sources are plotted in Fig.~\ref{fig:apparentlc}. The initial rise to the peak is not observed in the light curves shown, as the follow-up for \sniip\ started nearly $\sim\,$8 days after the predicted date of the explosion.

The $U$-band light curve has the steepest decline of all in the plateau phase, with the light curve declining almost linearly and the slope of which reduces after $\sim\,$45 days from the date of the explosion. The $B$-band light curve shows a relatively steeper decline in the initial part of the plateau phase, which flattens during the later stages (45--95 d). At the end of the plateau phase ($\sim\,$95 d), light curves in all the bands (except in $U$) decline steeply and settle onto a slowly declining radioactive decay-powered phase ($\sim\,$115 d). The $VRIgri$ light curves show a slow decline throughout the plateau phase which extends up to $\sim\,$95 days from the date of the explosion.

In order to compare the light curves of \sniip\ with other well-studied type II SNe, light curve parameters need to be extracted. Some type II SNe exhibit a change in the slope of the light curve at around 20-30 days from maximum, because the photosphere starts to recede deeper than the outer few tenths of a solar mass \citep{2010dessart}. To account for this, \citet{2014anderson} introduced the following parameters to characterize the light curves of type II SNe:\\ \\
(i) $s1$: the decline rate of the initial, steeper part of the plateau phase.\\
(ii) $s2$: the decline rate of the second, shallower part of the plateau phase.\\
(iii) $s3$: the decline rate in the radioactive tail phase.\\

The parameters $s1$, $s2$ and $s3$ are measured in units of magnitude per 100 days. We used Weighted Least-Square (WLS) optimization to fit different phases in the light curve with a straight line and extracted the decline rates in individual bands. The data points were weighted inversely with their respective error bars. As can be seen from the broadband light curves, \sniip\ shows a slight change in slope through its plateau ($s1\,>\,s2$). The light curve parameters $s1$, $s2$ and $s3$ in $UBVRI$ bands for \sniip\ are mentioned in Table~\ref{tab:lcpar}. The location of \sniip\ in the plot between the plateau decline rate ($s2$) and the peak $V$-band absolute magnitude ($\rm M_V^{max}$) is shown in Fig.~\ref{fig:mmaxs2}. The $B$-band decline rate, $\beta^{B}_{100}\,=\,3.13$ mag suggests that \sniip\ is a type II-P event \citep{1994patat}.

For comparison of \sniip\ with other type II SNe from the literature, we chose: normal type II-P SN 1999em \citep{2002aleonard} and SN 1999gi \citep{2002bleonard}, luminous and slow-declining type II-P SN 2004et \citep{2006sahu}, intermediate-luminosity type II-P SN 2012aw \citep{2013bose}, luminous and long-plateau type II-P SN 2009bw \citep{2012inserra}, luminous and fast-declining type II-P SN 2013ab \citep{2015boseab} and the luminous and fast-declining type II-P/L SN 2013ej \citep{2015boseej}, to see where \sniip\ lies amidst the class of type II SNe.

\begin{figure}
\centering
\resizebox{\hsize}{!}{\includegraphics{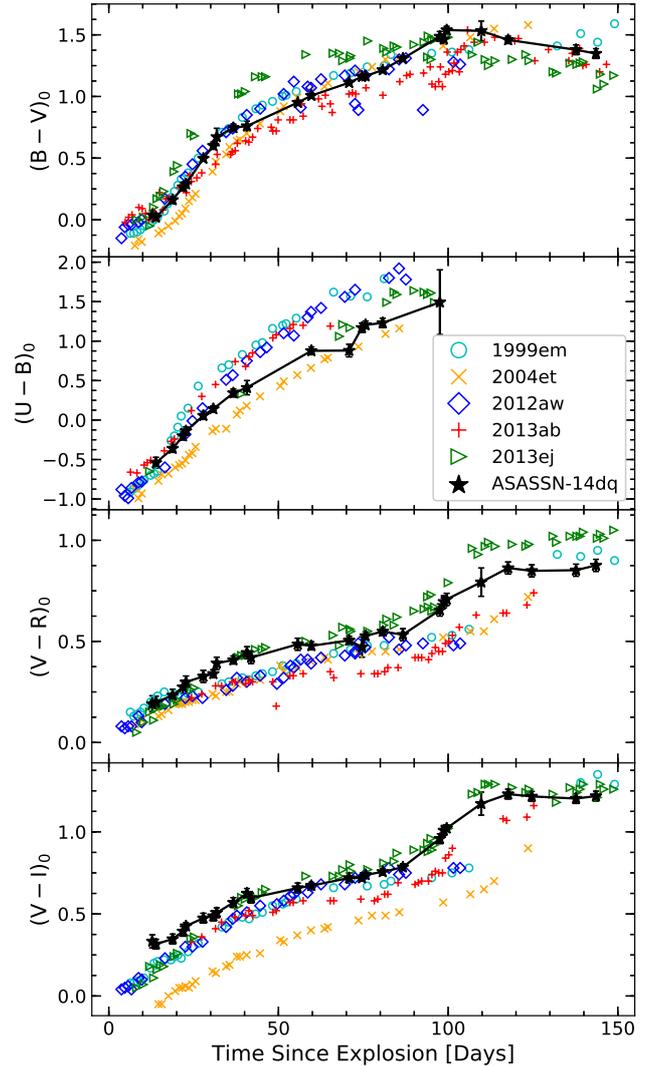}}
\caption{Evolution of intrinsic colour terms for \sniip\ along with other well-studied type II SNe. The colour terms for all SNe have been corrected for total (Galactic + host galaxy) extinction. The data for \sniip\ has been obtained from HCT and from \citet{2016valenti}. References for our sample of comparison is listed in Section~\ref{sec:applc}.}
\label{fig:colour}
\end{figure}

Evolution of intrinsic colours $B-V$, $U-B$, $V-R$ and $V-I$ of \sniip\ are shown in Fig.~\ref{fig:colour}. The temporal evolution of the SN envelope can be studied from these colour evolution plots. For comparison, we have plotted the colour evolution of \sniip\ with other well-studied type II SNe: SN 1999em, SN 2004et, SN 2012aw, SN 2013ab and SN 2013ej. The rapid decline of $U$ and $B$ fluxes in the early plateau phase leads to a rapid increase in the $U-B$ and $B-V$ colours, whereas the other colours rise slowly. The red-ward evolution of all the colours until the end of the plateau phase signifies the cooling of the ejecta. In all, the colour evolution of \sniip\ is similar to other type II SNe with no observed peculiarity.

Recently, a study on observed colours of type II SNe was carried out by \citet{2018dejaeger}. They inferred that the differences in observed colour is mostly intrinsic in origin and can be attributed to differences in progenitor radii and/or the presence/absence of CSM. They also found that fast-declining type II SNe have redder colours at later epochs (50--70 d) and attributed it to the interaction of the SN ejecta with the CSM. The colour evolution of \sniip\ in comparison with other type II SNe in our study appears to follow this trend.

%------------------------------------------------------------------------------%

\subsection{Absolute $V$-band magnitude and Bolometric light curve}

\begin{figure}
\centering
\resizebox{\hsize}{!}{\includegraphics{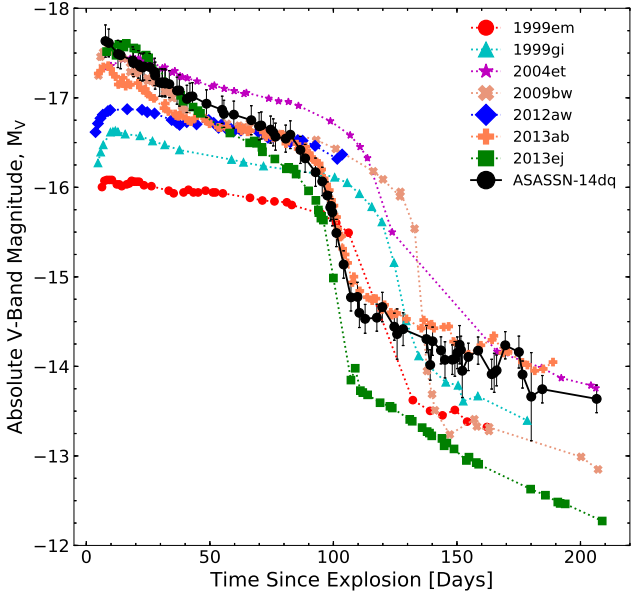}}
\caption{Absolute $V$-band magnitude light curve of \sniip\ along with other well studied type II SNe. References for our sample of comparison is listed in Section~\ref{sec:applc}.}
\label{fig:vabslc}
\end{figure}

Absolute broadband magnitudes were obtained from apparent magnitudes after correcting for reddening and distance estimated in the Sections~\ref{sec:losext} and~\ref{sec:dist}. The apparent SN magnitudes were corrected for extinction using the \citet{1999fitzpatrick} extinction law. The extinction coefficients in individual photometric bands were obtained using $R_{\lambda}$ corresponding to the central wavelength of the band (for $R_V\,=\,3.1$) and an \ebv\,=\,0.06 mag. The extinction corrected magnitudes were then converted to absolute magnitudes using a distance modulus, $\mu\,$=\,33.25$\,\pm\,$0.15 mag (equivalent to 44.8$\,\pm\,$3.1 Mpc).

$V$-band absolute magnitude light curve of \sniip\ is plotted along with our comparison sample in Fig.~\ref{fig:vabslc}. The early phase ($<\,$45 d) light curve of \sniip\ is slightly steep but flattens during the late-stages ($>\,$45 d) of the plateau phase. The $V$-band light-curve of \sniip\ in the nebular phase has a decline rate of 0.96 \maghundred\ and shows similarity to SN 2013ab and the decay rate of $\rm ^{56}Co$. The plateau to nebular phase transition occurs around $+$95 d and has a decline of $\sim\,$1.5 mag with a slope of 0.1 \magday. The peak $V$-band magnitude of \sniip, $\rm M_{V}$\,=\,--17.7$\,\pm\,$0.2 is much brighter than an average type II-P SNe \citep[which have an average peak $\rm M_V$\,=\,--15.66 mag]{2011li} and is among the brightest type II-P SN observed \citep{2011li, 2014anderson}. The mid-plateau $V$-band absolute magnitude ($\rm M_{V}^{50}$\,=\,--16.9$\,\pm\,$0.2 mag) and the plateau duration ($\sim\,$90 d) of \sniip\ is comparable to SN 2013ej and SN 2013ab. 

\begin{figure}
\centering
\resizebox{\hsize}{!}{\includegraphics{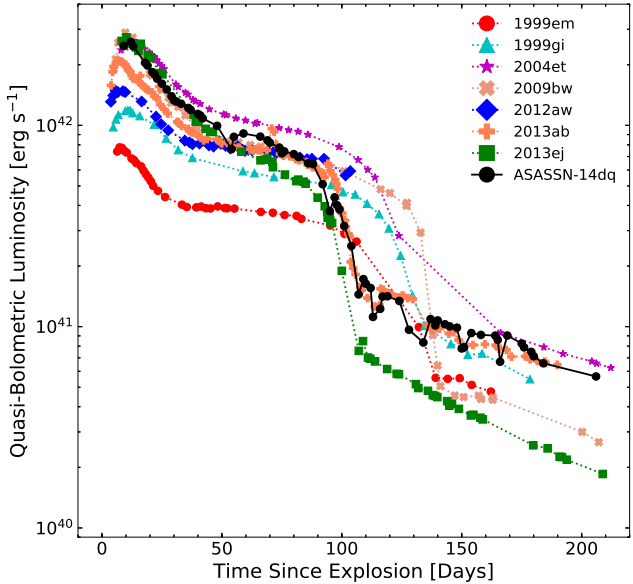}}
\caption{Quasi-Bolometric light curve of \sniip\ along with other well studied type II SNe. References for our sample of comparison is listed in Section~\ref{sec:applc}.}
\label{fig:bolometriclc}
\end{figure}

Quasi-bolometric light curve for \sniip\ was obtained using $UBVRI$ data obtained from HCT and $BVgri$ data from \citet{2016valenti}. The extinction corrected SN magnitudes were converted to monochromatic fluxes with the help of the following relation from \citet{1998bessell}:
\begin{equation}
 \rm m_{\lambda} = -2.5 * log(f_{\lambda}) - 21.100 - zp(f_{\lambda}),
\end{equation}

where $zp(f_{\lambda})$ is the wavelength dependent zero point magnitude based on $\rm M_V$\,=\,0.03 mag for Vega. The zero points for Bessell's $UBVRI$ filters were obtained from \citet{1998bessell}, and for SDSS $gri$ filters from the SVO Filter Profile Service website\footnote{\url{http://svo2.cab.inta-csic.es/theory/fps/}}.

The quasi-bolometric flux for an epoch was then calculated by integrating over a cubic spline fit to the $UBVRIgri$ photometric fluxes at central wavelengths of the respective band-pass filters. The flux was then integrated over the wavelength range of 3100 to 9200 \AA. The wavelength range was chosen as the edges (1\% of maximum transmission) of the transmission function of the $U$ and $I$ bands. The quasi-bolometric flux of the sample used for comparison was also calculated by integrating over the same wavelength range. For the missing band-pass values on some epochs, the magnitude of the band-pass filter was obtained by interpolating the light curve using a spline. The quasi-bolometric luminosity (hereafter bolometric) was obtained after correcting the flux obtained for the distance to the host galaxy, D\,=\,$44.8\pm3.1$ Mpc (see Section~\ref{sec:dist}).

The bolometric light curve of \sniip\ is shown in Fig.~\ref{fig:bolometriclc} along with the bolometric light curves of our comparison sample. During the early parts of the plateau phase ($<$\,45 d), the bolometric flux is dominated by the flux from the shorter wavelength bands \textit{i.e} $UB$, which is reflected in the bolometric LC as the steep decline phase of the plateau. During the later stages of the plateau phase ($>$\,45 d), the major contribution to the bolometric flux shifts towards the $VRI$ bands and the UV contribution becomes insignificant as the ejecta had cooled with time.

The peak bolometric luminosity of \sniip\ is $\rm \sim2.6\times10^{42} erg\ s^{-1}$ which classifies it as a luminous type II-P SN. The bolometric luminosity declines at the rate of 0.36 dex $\rm (100\ d)^{-1}$ during the plateau phase and 0.15 dex $\rm (100\ d)^{-1}$ during the radioactive decay-powered phase.

%------------------------------------------------------------------------------%
\section{Spectroscopic Evolution}
\label{sec:specevol}
%------------------------------------------------------------------------------%

The spectral evolution of \sniip\ over 23 epochs is displayed in Figs.~\ref{fig:specplat} and~\ref{fig:specneb} for the plateau and nebular phase, respectively. All the spectra were corrected for the recession velocity of the host galaxy (\hostglx) using the value of redshift (z\,=\,0.010424) obtained from NED. This redshift is consistent with the value obtained (z\,=\,0.010) by Supernova Identification software (SNID\footnote{\url{https://people.lam.fr/blondin.stephane/software/snid/}}) \citep{2007blondin}. Identification of spectral features in \sniip\ was done as in prior studies of type II-P SNe \citep{2006sahu, 2015boseab} and is displayed in Fig.~\ref{fig:specline}.

\subsection{Early phase spectra}

\begin{figure*}
\centering
\resizebox{\hsize}{!}{\includegraphics{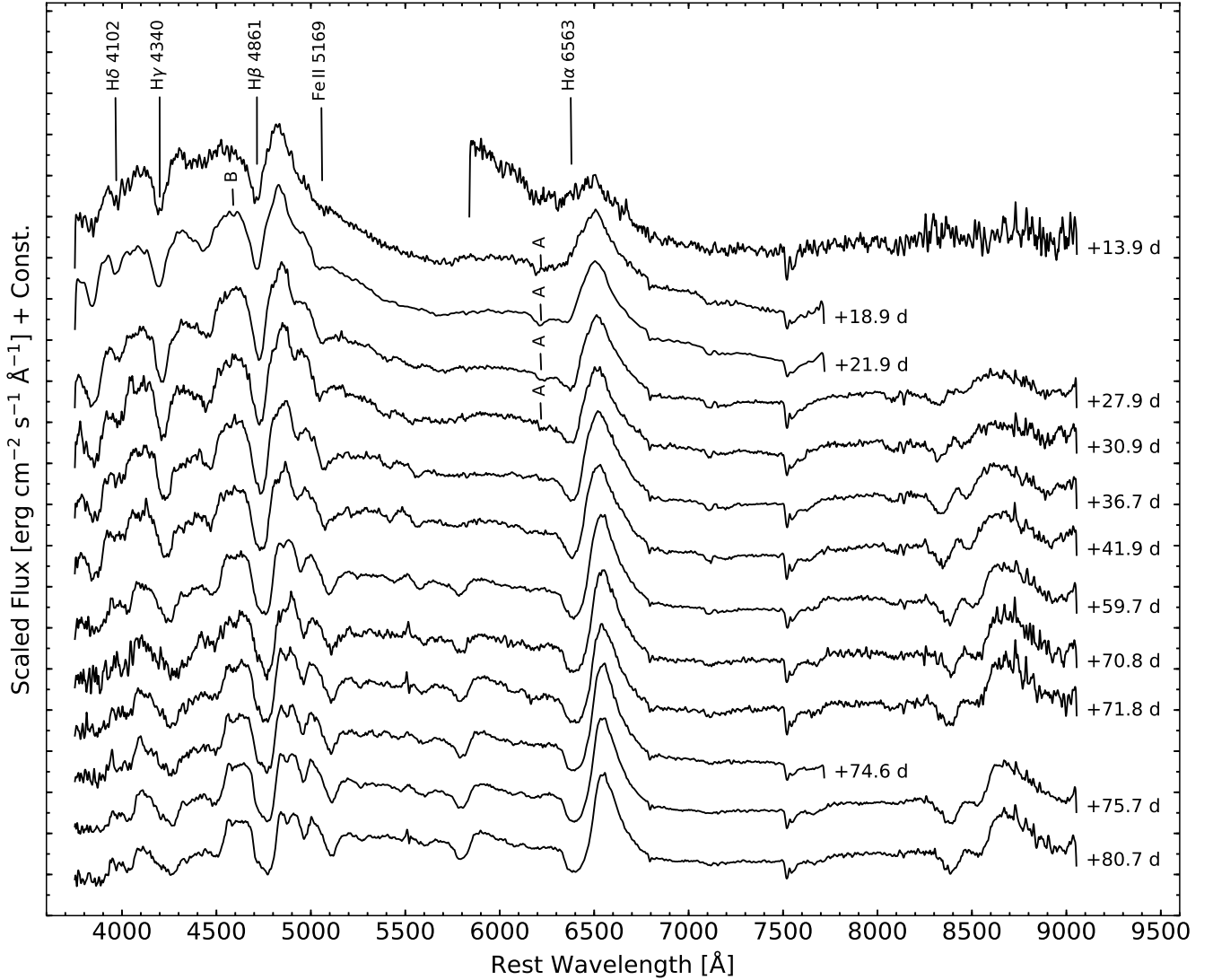}}
\caption{Spectroscopic evolution of \sniip\ in the plateau phase (till $\sim90$ days since the date of explosion). The P-Cygni profiles of hydrogen Balmer (H$\alpha$, H$\beta$ etc.) lines can be seen in the spectra.}
\label{fig:specplat}
\end{figure*}

\begin{figure*}
\centering
\resizebox{\hsize}{!}{\includegraphics{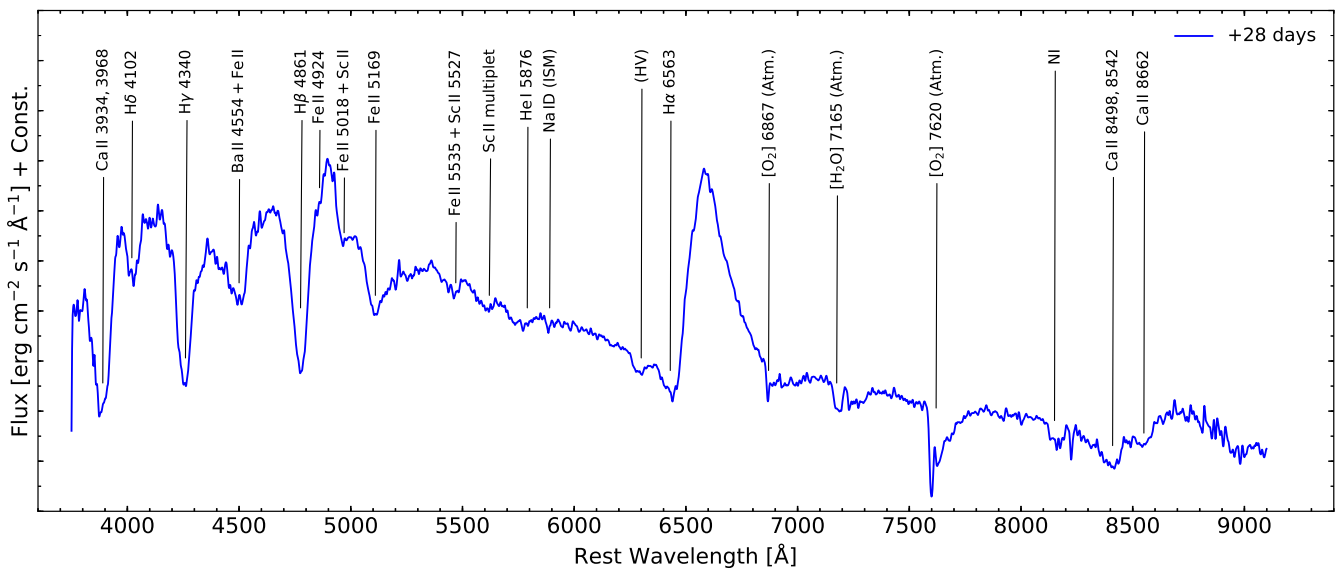}}
\caption{Line identification of spectral features in \sniip\ in the plateau phase.}
\label{fig:specline}
\end{figure*}

\begin{figure}
\centering
\resizebox{\hsize}{!}{\includegraphics{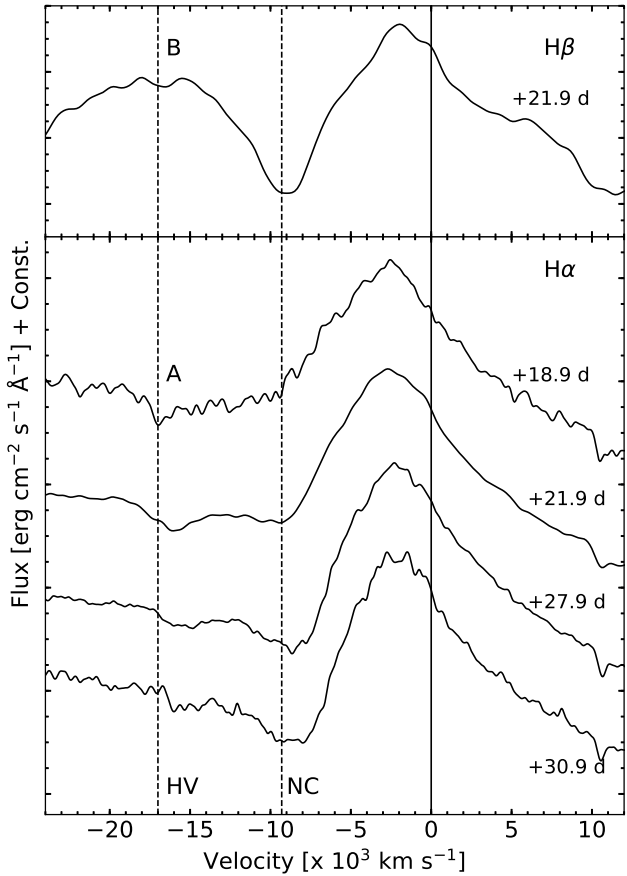}}
\caption{P-Cygni profiles of H$\beta$ (top panel) and H$\alpha$ (bottom panel) of \sniip. The high-velocity (HV) feature of H$\alpha$ is initially seen at $\sim\,$17000 \kms\ along with the slower normal component (NC) in the spectrum of $\sim\,$19 d. The H$\beta$ counterpart to the HV feature of H$\alpha$ is also seen in the spectrum on $\sim\,$22 d. The solid line denotes the zero velocity of the P-Cygni features with the dashed lines indicating the NC and the HV feature.}
\label{fig:hv}
\end{figure}

The early phase spectra of a SN help determine the ejecta composition and the nature of the progenitor as the photosphere recedes the surface exposing the inner layers. The earliest spectrum of \sniip\ obtained $\sim\,$14 days from the explosion is quite noisy and displays a developed emission-dominated P-Cygni profile of H$\alpha$. The blue-shifted minimum of the absorption feature of H$\alpha$ gave an expansion velocity of $\sim\,$11500 \kms. Fe\,II 5169 \AA\ starts appearing early ($\sim\,$22 d) in the plateau along with Ba\,II 4554 \AA\ blend. Weaker lines of Fe\,II 4924, 5018 \AA\ and Ca\,II IR triplet 8498, 8542, 8602 \AA\ start appearing around $\sim\,$28 d and are traceable in the spectrum of $\sim\,$31 d. The emergence of lines from heavier atoms like iron, calcium, scandium, barium etc. around $\sim\,$22 d signify essentially the composition of the progenitor star as the photosphere traces the outer parts of the ejecta up to $\sim\,$80 days. As the SN evolves further through the plateau phase, H$\alpha$, H$\beta$, Ca\,II IR triplet absorption features start becoming narrower and deeper due to the decreasing ejecta velocity. Weak signatures of Na\,I D interstellar absorption features from the MW can also be seen in the spectra having a high SNR (i.e. $\sim\,$28 d). Signatures of Ca\,II doublet 3934, 3968 \AA\ can be seen throughout the plateau phase.

An absorption is seen blue-ward of the H$\alpha$ component in the spectra obtained during $\sim\,$19--31 days (see Fig.~\ref{fig:specplat}). This could be due to Si\,II $\lambda$6355, or could be a high-velocity (HV) component of H$\alpha$ \citep[see description on \lq \lq Cachito\rq \rq\ in][]{2017gutierrez}. The feature (marked \lq A\rq) seen in the spectrum of $\sim\,$19 d is observed at $\sim\,$17000 \kms\ \textit{w.r.t.} the H$\alpha$ rest wavelength and is significantly higher than the normal component ($\sim\,$9500 \kms) of H$\alpha$ at that epoch. The spectrum on $\sim\,$22 d shows an absorption blue-ward of H$\beta$, marked as feature \lq B\rq\ in Fig.~\ref{fig:specplat}. If this absorption is associated with H$\beta$, it has a velocity similar to the absorption feature \lq A\rq, supporting the identity of these features to be the high-velocity (HV) components of hydrogen \citep[see][]{2012inserra, 2017gutierrez}. The P-Cygni profiles of H$\alpha$ and H$\beta$ are plotted in Fig.~\ref{fig:hv} for the period over which these HV features are seen. As the feature \lq A\rq doesn't follow the usual red-ward drift (slow evolution compared to the normal component) in time, it is highly unlikely to be the Si\,II $\lambda$6355 from the SN \citep{2014afaran}. This two-component P-Cygni profile of H$\alpha$ can hence signify the existence of high-velocity (HV) material in the outer envelope of \sniip\ in the early phases. This feature was also present in the spectra of SN 2004et \citep{2006sahu} and SN 1999em \citep{2002bleonard} and could arise from the interaction of the SN ejecta with the CSM surrounding the pre-supernova star \citep{2007chugai}. The HV component is initially stronger than the normal component but fades away \textit{w.r.t.} the normal component as the SN enters the mid-plateau ($\sim\,$42 d) phase.

Weak signature of He\,I 5876 can be seen in the spectrum of $\sim\,$28 d. Na\,I D from the SN starts appearing at a similar location in the spectrum obtained on $\sim\,$37 d and the He\,I feature is no longer discernible \citep{2017gutierrez}. The Fe\,II 5535 \AA\ and Sc\,II 5527 \AA\ blend along with Sc\,II 5665 \AA\ multiplet start appearing in the spectra by $\sim\,$28 d and are distinguishable by $\sim\,$42 d. As the supernova ages across the plateau, absorption features from numerous metal lines become narrower and deeper. Ba\,II 6142 \AA\ and Sc\,II 6246 \AA\ are seen in the spectrum taken at $\sim\,$45 d. The spectrum of \sniip\ at $\sim\,$80 d still shows strong absorption features signifying that the SN has not yet entered the nebular phase. The above-mentioned features are present in the spectra until $\sim\,$104 d and their comparison with other SNe (top two sub-plots of Fig.~\ref{fig:speccomp}) indicates that the plateau phase spectral features are similar (although weaker) to other type II-P events in comparison.

\subsection{Nebular phase spectra}

\begin{figure*}
\centering
\resizebox{\hsize}{!}{\includegraphics{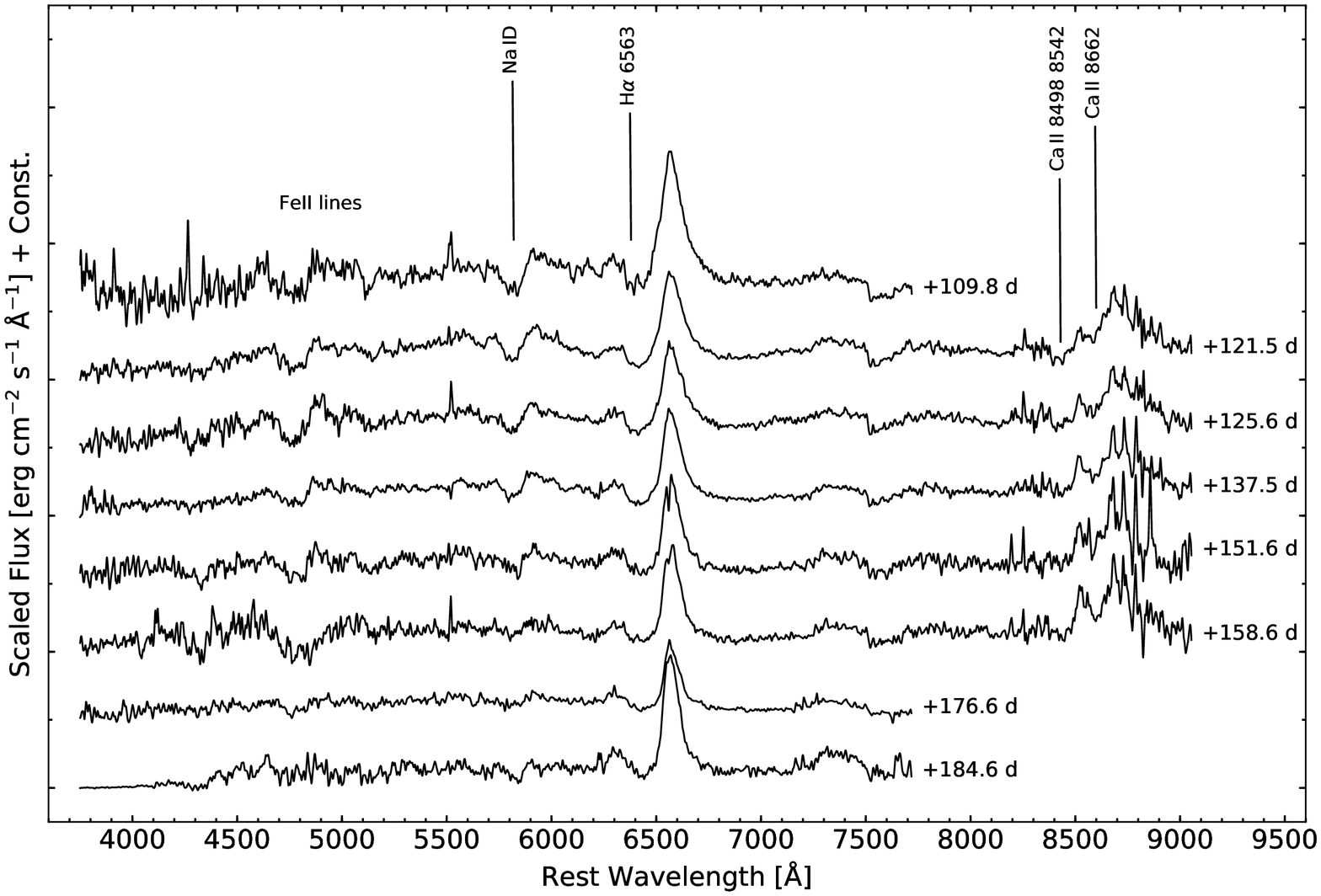}}
\caption{Spectroscopic evolution of \sniip\ in the nebular phase (started $\sim\,$105 days from the date of explosion). P-Cygni profiles of hydrogen can still be seen in the spectra with shallow absorption troughs. The spectra in this phase have a flat continuum.}
\label{fig:specneb}
\end{figure*}

\begin{figure}
\centering
\resizebox{\hsize}{!}{\includegraphics{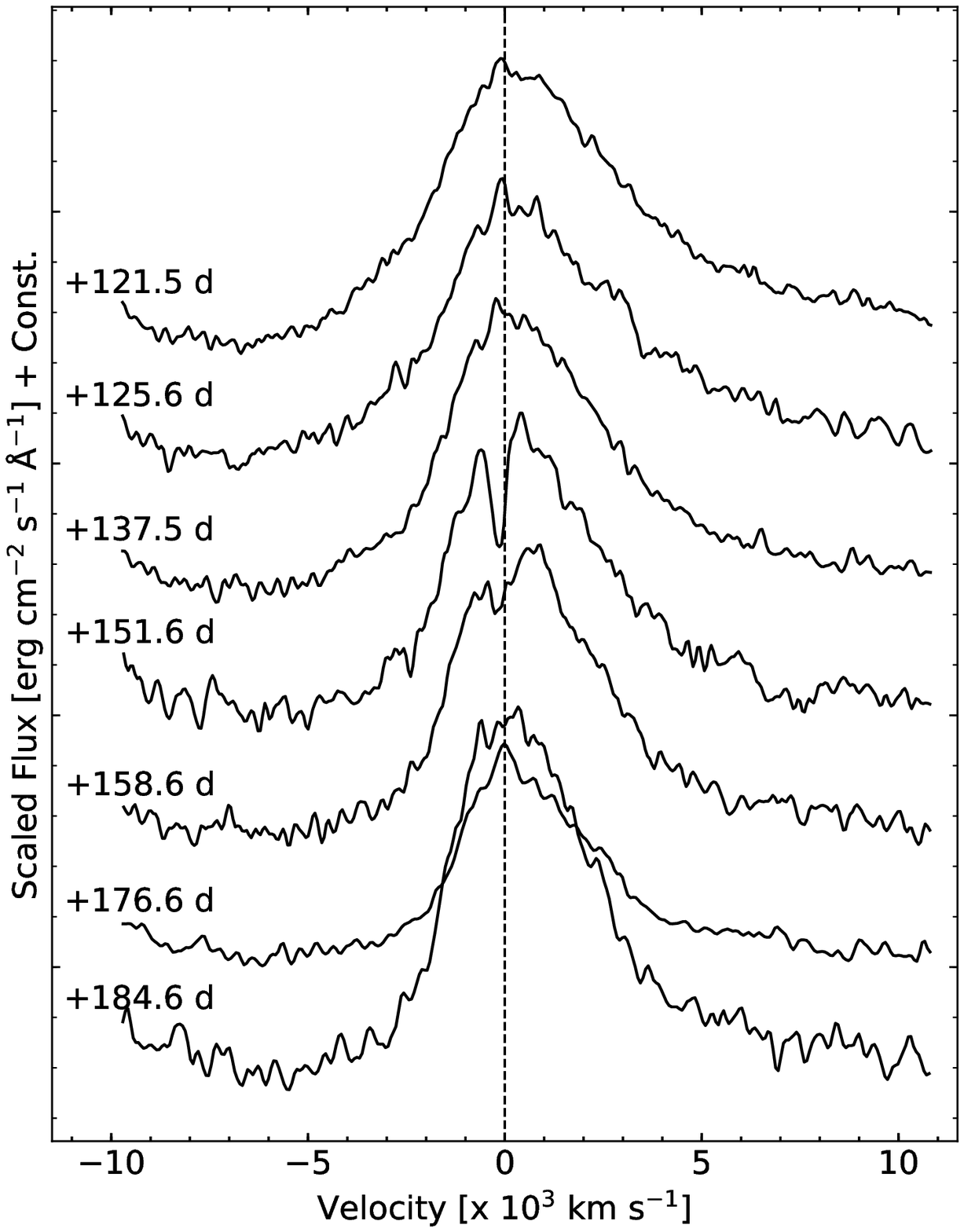}}
\caption{Evolution of the H$\alpha$ P-Cygni profile in the nebular phase of \sniip. A \lq \lq notch\rq \rq\ can be seen in the spectra after $\sim\,$126 d. This asymmetry in emission profile of H$\alpha$ signify CSM interaction of the ejecta.}
\label{fig:asymmetry}
\end{figure}

The nebular phase in a supernova is marked by the presence of strong emission features in its spectra which is attributed to its optically thin ejecta. The spectrum mostly forms deep inside the ejecta, revealing information on the nucleosynthesis that occurred during the explosion of the star. The spectra of \sniip\ in the nebular phase is plotted in the Fig.~\ref{fig:specneb}. The spectrum of \sniip\ at $\sim\,$120 d shows a flat continuum. Na\,I D, H$\alpha$ and Ca\,II triplet 8498, 8542, 8662 \AA\ show shallow absorption troughs in the early nebular phase which fades away with time. These features also become progressively emission dominated as the supernova ages. The emission profile of H$\alpha$ in this phase shows an asymmetric feature beginning $\sim\,$126 d and is highlighted in Fig.~\ref{fig:asymmetry}. Forbidden lines from [O\,I] 6300, 6364 \AA\ and [Ca\,II] 7291, 7324 \AA\ can be seen in the spectra taken after $\sim\,$122 d, which strengthen as the supernova progresses in the nebular phase. The Ca\,II triplet becomes comparable to H$\alpha$ in terms of line luminosity in the nebular phase ($\sim\,$159 d), although the [O\,I] doublet stays quite weak in \sniip.

%------------------------------------------------------------------------------%

\subsection{Expansion Velocity}

\begin{figure}
\centering
\resizebox{\hsize}{!}{\includegraphics{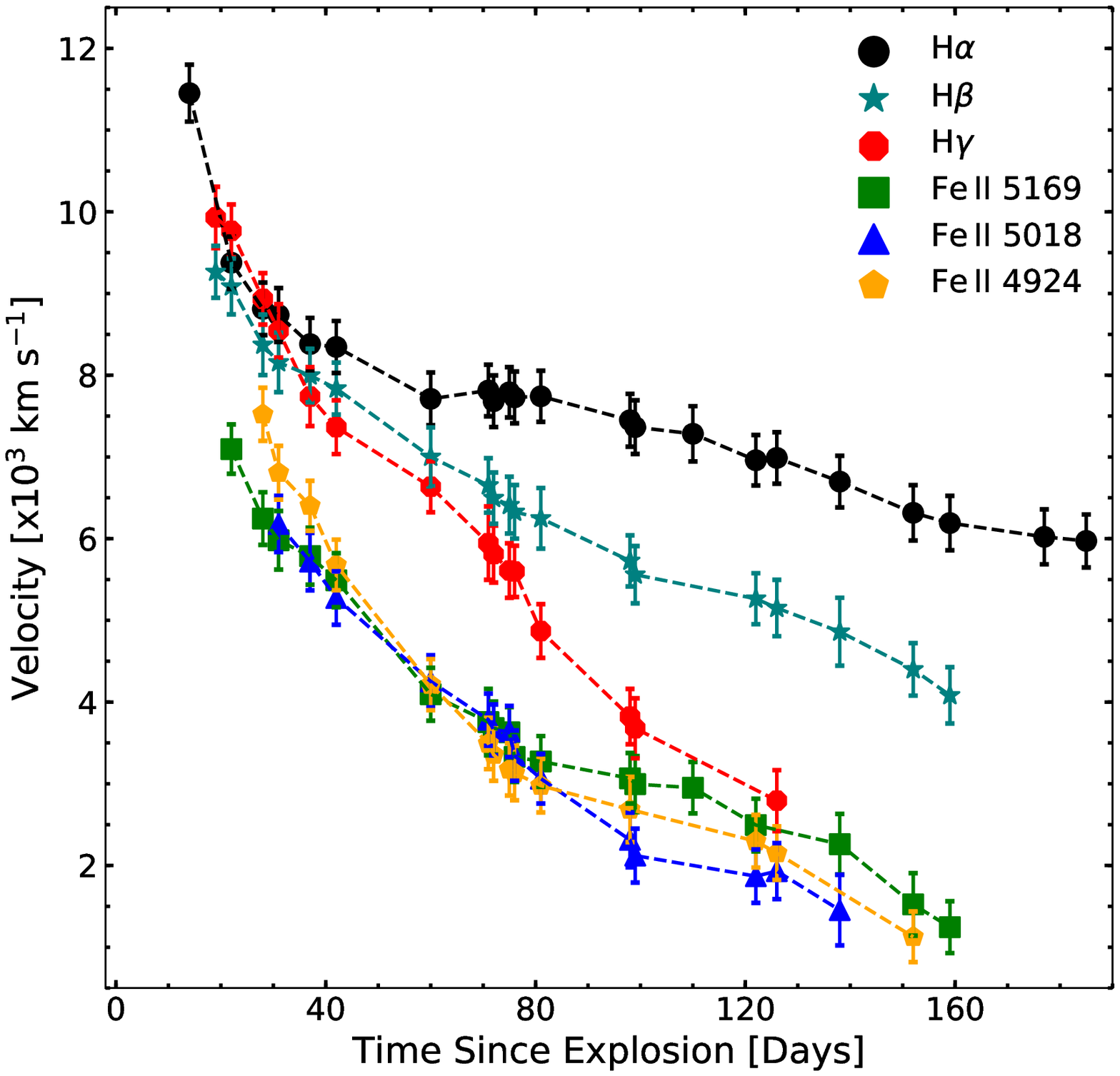}}
\caption{Velocity evolution of spectral features: H$\alpha$, H$\beta$, H$\gamma$ and Fe\,II triplet. The velocities are determined using the blue-shifted absorption minima of the P-Cygni profile.}
\label{fig:photvel}
\end{figure}

Spectral features originating from different layers in the ejecta have different characteristic velocities and hence can be used to study the geometry of the explosion. They also offer information on the energetics of the explosion. Photosphere of a type II-P SN refers to the ionized, thin spherical shell which radiates most of the continuum radiation as a \lq\lq dilute blackbody\rq\rq\ \citep{2002bleonard}. The continuum spectrum released from the electron-scattering photosphere (defined as having an optical depth of photons equal to $\sim\,$2/3) is actually produced in a deeper layer at which the radiation field thermalizes at the temperature of the local gas. In Fig.~\ref{fig:photvel}, we plotted the line velocity evolution of spectral features H$\alpha$, H$\beta$, H$\gamma$, and Fe\,II 4924, 5018, 5169 \AA. The line velocities were calculated from the minimum of the absorption features determined by fitting a Gaussian profile to the same using IRAF in the redshift corrected spectra. The error in estimating velocity includes the error in the measurement along with the uncertainty in the wavelength scale, as mentioned in \citet{2002aleonard}.

Due to the mixing of layers in the SN ejecta caused by the shock breakout and the inward movement of the recombination front (in mass coordinate), a single spectral line cannot truly depict the true photospheric velocity throughout the course of evolution of the SN. Balmer lines present in the spectra of a SN can represent the photospheric velocity fairly accurately only when the optical depths of these lines are low. This is true for spectra taken during the very early phase ($\leq$\,8 d) of a type II-P SN. As the SN fades away with time, velocities determined from the hydrogen Balmer lines become a poor representation of the photospheric velocity \citep{1989eastman}. These features yield a velocity higher than the true photospheric velocity because most of the absorption happens in the outer (faster) layers of the SN ejecta. During the late phase, weak and unblended absorption features of metals (Fe, Sc etc.) can be used to determine the photospheric velocity \citep{2001hamuy}. Hence, Fe\,II 4924, 5018, 5169 \AA\ lines were used to determine the photospheric velocity during the plateau phase. The velocities inferred from metal lines follow a similar trend and form in deeper (slower) layers compared to the hydrogen Balmer lines which form in outer (faster) layers.

The photospheric velocity inferred for \sniip\ varies from $\sim\,$11500 \kms\ in the earliest spectra ($\sim\,$14 d) to $\sim\,$1200 \kms\ in the spectra obtained on $\sim\,$160 d. It is to be noted here that, due to the possible blending of Fe\,II 5169 \AA\ with other metal lines after the end of the plateau phase ($>$\,95 d), Fe\,II 5169 \AA\ absorption feature may not reflect the photospheric velocity accurately. Also, the velocity estimated from Fe\,II 4924 \AA\ feature might display consistently lower values because of its suspected blending with the Ba\,II 4934 \AA\ feature \citep{2005hendry}. However, in the case of our low-resolution spectra, we are unable to discern such differences because the velocities determined from the individual Fe\,II features lie within 1$\sigma$ error of each other.

%-----------------------------------------------------------------------------%
\section{Nickel Mass}
\label{sec:nickelmass}
%-----------------------------------------------------------------------------%

CCSNe produce radioactive $\rm ^{56}Ni$ through the explosive nucleo-synthesis of Si \citep{1980arnett}. The nebular phase light curve in type II SNe is mainly powered by the radioactive decay of $\rm ^{56}Ni \rightarrow\ ^{56}Co \rightarrow\ ^{56}Fe$. The $\gamma$-ray photons and positrons emitted from the above decay thermalize the ejecta and power the tail of the light curve.

\subsection{Rediscovering the correlation between Nickel mass and Steepness parameter}
\label{sec:nickelsteepness}

\begin{figure*}
\centering
\resizebox{\hsize}{!}{\includegraphics{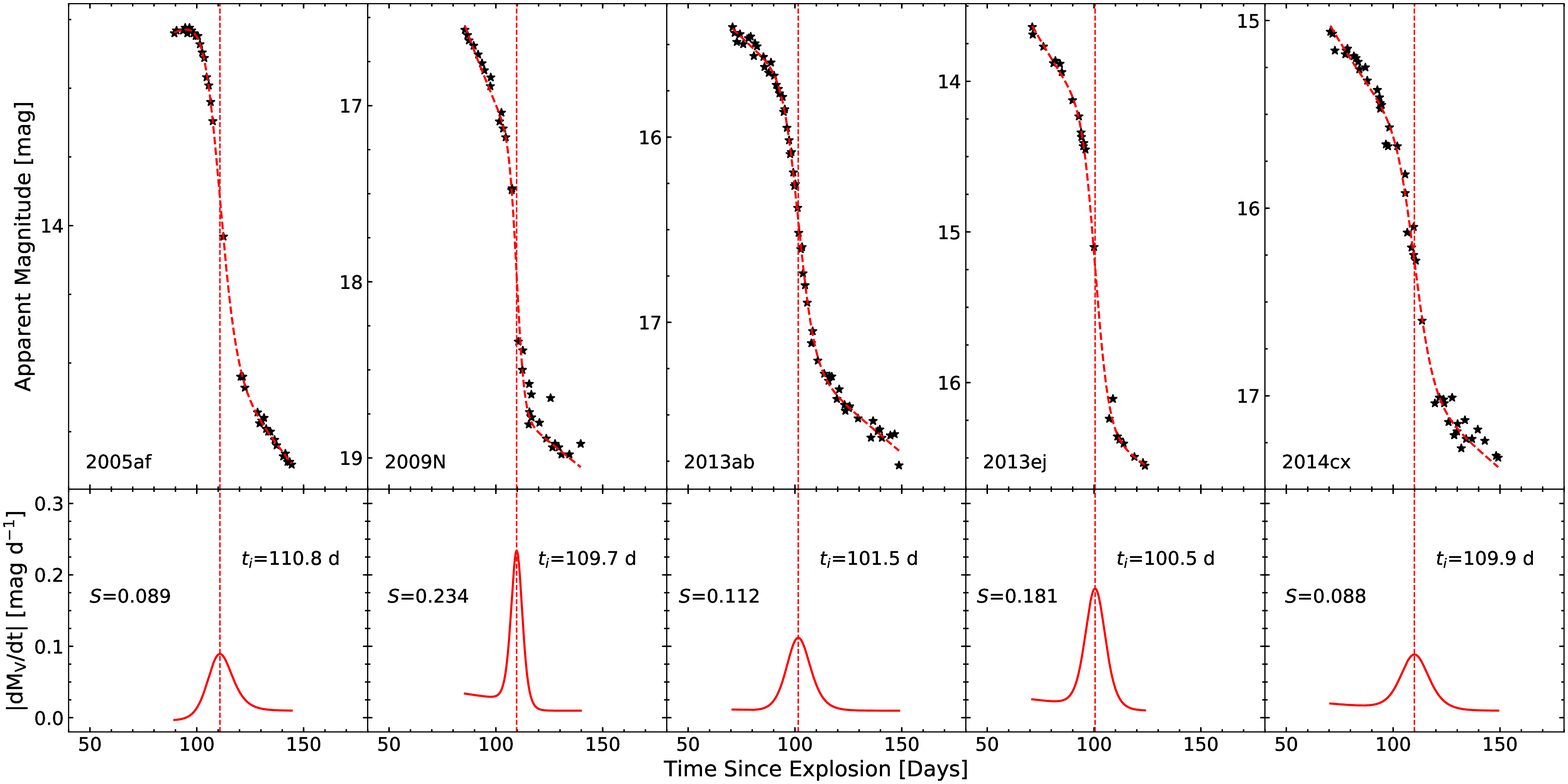}}
\caption{Steepness parameter $S$ and point of inflection $t_i$ determined for type II SNe: 2005af, 2009N, 2013ab, 2013ej and 2014cx. The \textit{upper} panel in the plot displays the apparent $V$-band light curve (starred) along with the best fit (dashed). The \textit{lower} panel shows steepness (the slope) of the light curve as a function of time. The steepness parameter and the point of inflection are also mentioned for each SN in the $lower$ panel. The vertical dashed line marks the epoch of inflection.}
\label{fig:steepness}
\end{figure*}

\begin{figure}
\centering
\resizebox{\hsize}{!}{\includegraphics{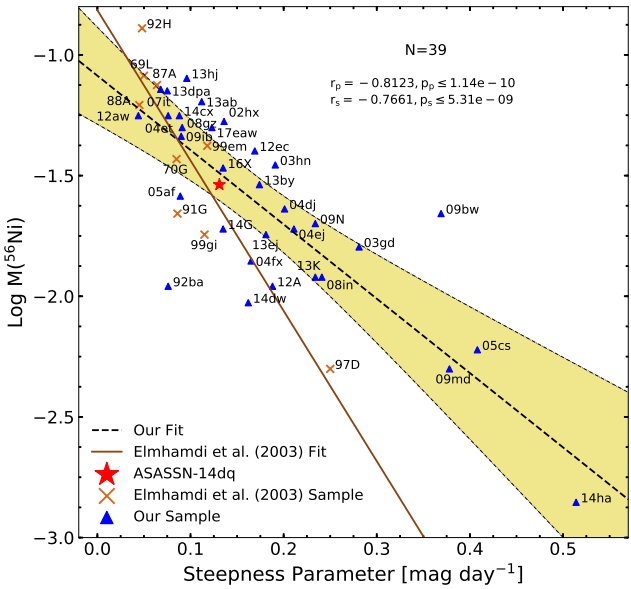}}
\caption{Correlation between $\rm\ Log\ M(^{56}Ni)$ and Steepness parameter $S$. The number of SNe, Pearson (suffix $p$) and Spearman (suffix $s$) correlation coefficients along with the chance probability of finding a correlation is mentioned in the above figure. The 3$\sigma$ confidence interval of the fit has been shown shaded in $dark-yellow$ colour.}
\label{fig:nickelsteepness}
\end{figure}

\begin{table*}
\centering
\caption{Estimates of Steepness $S$, time of inflection $t_i$ and $\rm ^{56}Ni$ mass for 39 type II SNe.}
\setlength{\tabcolsep}{4pt}
\label{tab:lc_steepness}                     
\begin{tabular}{l c c c c c c}          
\hline
\hline                      
SN      &  Steepness        &   $\rm t_{i}$         &   Reference       &   $\rm M_{Ni}\ (M_{\odot}$)   &   $\rm M_{Ni}$ ($\rm M_{\odot}$)  &   Reference           \\
(Name)  &  (\magday)        &   (d)                 &   (Data)          &    (From Steepness, Ref. 1)   &   (From Literature)               &  ($\rm ^{56}Ni$ Mass) \\
\noalign{\smallskip} \hline \noalign{\smallskip}
 & & & \citet{2003elmhamdi} sample & & & \\
\noalign{\smallskip} \hline \noalign{\smallskip}                  
1969L   &   0.050           &   110$\,\pm\,$4       &     2             &      0.064$\,\pm\,$0.001        &   0.082$^{+0.034}_{-0.026}$   &   3        \\
1970G   &   0.085           &    94$\,\pm\,$4       &     2             &      0.049$\,\pm\,$0.001        &   0.037$^{+0.019}_{-0.012}$   &   3        \\
1987A   &   0.064           &   107$\,\pm\,$2       &     2             &      0.057$\,\pm\,$0.001        &   0.075                       &   3        \\
1988A   &   0.045           &   134$\,\pm\,$4       &     2             &      0.067$\,\pm\,$0.001        &   0.062$^{+0.029}_{-0.020}$   &   3        \\
1991G   &   0.086           &   122$\,\pm\,$5       &     2             &      0.048$\,\pm\,$0.001        &   0.022$^{+0.008}_{-0.006}$   &   3        \\
1992H   &   0.048           &   112$\,\pm\,$5       &     2             &      0.065$\,\pm\,$0.001        &   0.129$^{+0.053}_{-0.037}$   &   3        \\
1997D   &   0.250           &   112$\,\pm\,$5       &     2             &      0.013$\,\pm\,$0.001        &   0.005$\,\pm\,$0.004             &   4        \\
1999em  &   0.118           &   112$\,\pm\,$5       &     2             &      0.037$\,\pm\,$0.001        &   0.042$^{+0.027}_{-0.019}$   &   3        \\
1999gi  &   0.115           &   120$\,\pm\,$3       &     2             &      0.038$\,\pm\,$0.001        &   0.018$^{+0.013}_{-0.009}$   &   3        \\
\noalign{\smallskip} \hline \noalign{\smallskip}
 & & & Additional sample used in this paper & & & \\
\noalign{\smallskip} \hline \noalign{\smallskip}
2002hx  &   0.136           &    75.7               &      5, 27        &      0.032$\,\pm\,$0.001        &   0.053$^{+0.016}_{-0.023}$   &   5        \\
2003gd  &   0.281           &   124.0               &      6            &      0.010$\,\pm\,$0.001        &   0.016$^{+0.010}_{-0.006}$   &   6        \\
2003hn  &   0.191           &    97.3               &      5, 27        &      0.021$\,\pm\,$0.001        &   0.035$^{+0.008}_{-0.011}$   &   5        \\
2004dj  &   0.201           &   126.2               &      7            &      0.019$\,\pm\,$0.001        &   0.023$\,\pm\,$0.005             &   7        \\
2004ej  &   0.211           &   114.4               &      5            &      0.018$\,\pm\,$0.001        &   0.019$^{+0.005}_{-0.007}$   &   5        \\
2004et  &   0.076           &   123.1               &      8            &      0.052$\,\pm\,$0.001        &   0.056$\,\pm\,$0.040             &   4        \\
2004fx  &   0.165           &   104.6               &      5            &      0.025$\,\pm\,$0.001        &   0.014$^{+0.004}_{-0.006}$   &   5        \\
2005af  &   0.089           &   110.8               &      5            &      0.047$\,\pm\,$0.001        &   0.026$^{+0.012}_{-0.021}$   &   5        \\   
2005cs  &   0.408           &   125.8               &      9            &      0.0036$\,\pm\,$0.0004      &   0.006$\,\pm\,$0.003             &   4        \\
2007it  &   0.068           &   108.7               &      12           &      0.056$\,\pm\,$0.001        &   0.072$^{+0.031}_{-0.054}$   &   12       \\   
2008gz  &   0.091           &   120.1               &      24           &      0.046$\,\pm\,$0.001        &   0.050$\,\pm\,$0.010             &   24       \\
2008in  &   0.241           &   107.8               &      10           &      0.014$\,\pm\,$0.001        &   0.012$\,\pm\,$0.005             &   4        \\
2009N   &   0.234           &   109.7               &      11           &      0.015$\,\pm\,$0.001        &   0.020$\,\pm\,$0.004             &   4        \\
2009ib  &   0.090           &   140.2               &      13           &      0.047$\,\pm\,$0.001        &   0.046$\,\pm\,$0.015             &   13       \\
2009md  &   0.378           &   118.9               &      25           &      0.0046$\,\pm\,$0.0004      &   0.005$\,\pm\,$0.001             &   4        \\
2012A   &   0.188           &   107.4               &      18           &      0.021$\,\pm\,$0.001        &   0.011$\,\pm\,$0.001             &   14       \\
2012aw  &   0.044           &   130.0               &      18           &      0.068$\,\pm\,$0.001        &   0.056$\,\pm\,$0.013             &   15       \\
2012ec  &   0.169           &   106.8               &      16           &      0.025$\,\pm\,$0.001        &   0.040$\,\pm\,$0.015             &   16       \\
2013K   &   0.234           &   128.2               &      22           &      0.015$\,\pm\,$0.001        &   0.012$\,\pm\,$0.010             &   22       \\
2013ab  &   0.112           &   101.5               &      17           &      0.039$\,\pm\,$0.001        &   0.064$\,\pm\,$0.006             &   17       \\
2013by  &   0.174           &    87.1               &      18           &      0.024$\,\pm\,$0.001        &   0.029$\,\pm\,$0.005             &   18       \\
2013ej  &   0.181           &   100.5               &      19           &      0.022$\,\pm\,$0.001        &   0.018$\,\pm\,$0.006             &   18       \\
LSQ13dpa &   0.075          &   126.5               &      5, 27        &      0.053$\,\pm\,$0.001        &   0.071$\,\pm\,$0.013             &   27       \\   
2013hj  &   0.096           &   106.4               &      28           &      0.044$\,\pm\,$0.001        &   0.080$\,\pm\,$0.008             &   28       \\
2014G   &   0.135           &    86.8               &      26           &      0.032$\,\pm\,$0.001        &   0.019$\,\pm\,$0.003             &   18       \\
2014cx  &   0.088           &   109.9               &      20, 27       &      0.047$\,\pm\,$0.001        &   0.056$\,\pm\,$0.008             &   20       \\
2014dw  &   0.162           &    91.3               &      27           &      0.026$\,\pm\,$0.001        &   0.0094$\,\pm\,$0.0008           &   5        \\
ASASSN-14ha &   0.514       &   136.8               &      27           &      0.0015$\,\pm\,$0.0002      &   0.0014$\,\pm\,$0.0002           &   27       \\
2016X   &   0.135           &    94.9               &      21           &      0.032$\,\pm\,$0.001        &   0.034$\,\pm\,$0.006             &   21       \\
2017eaw &   0.137           &   120.5               &      23           &      0.036$\,\pm\,$0.001        &   0.050$\,\pm\,$0.015             &   23       \\
\noalign{\smallskip} \hline \noalign{\smallskip} 
\end{tabular}
\newline
(1) This paper; 
(2) \citet{2003elmhamdi};
(3) \citet{2003hamuy};
(4) \citet{2014spiro};
(5) \citet{2014anderson}
(6) \citet{2005hendry};
(7) \citet{2006zhang};
(8) \citet{2006sahu};
(9) \citet{2009pastorello};
(10) \citet{2011roy};
(11) \citet{2014takats};
(12) \citet{2011andrews};
(13) \citet{2015takats};
(14) \citet{2013tomasella};
(15) \citet{2013bose};
(16) \citet{2015barbarino};
(17) \citet{2015boseab};
(18) \citet{2015valenti};
(19) \citet{2015boseej};
(20) \citet{2016huang};
(21) \citet{2018huang};
(22) \citet{2018tomasella};
(23) \citet{2018tsvetkov};
(24) \citet{2011broy};
(25) \citet{2011fraser};
(26) \citet{2016terreran};
(27) \citet{2016valenti};
(28) \citet{2016bose}.
\end{table*}

\citet{2003elmhamdi}, using a sample of 10 type II SNe determined that the $\rm ^{56}Ni$ mass anti-correlates with the maximum value of the steepness parameter, $S\,=\,-dM_V/dt$ during the transition phase. We explore this correlation further as this can help probe $\rm ^{56}Ni$ masses for type II SNe independent of their distance and extinction. To accommodate the growing sample of type II SNe, we rebuild the empirical relation after including 30 type II SNe \citep[in addition 9 SNe from][]{2003elmhamdi} having a good photometric coverage in the transition phase of the $V$-band light curve and known $\rm ^{56}Ni$ mass. In order to calculate $S$, we follow the procedure in \citet{2003elmhamdi}. Firstly, we obtain the $V$-band fluxes from magnitudes and fit the transition period from the plateau to the radioactive tail by a function comprising of terms representing the plateau and the radioactive-decay phase:

\begin{equation}
\label{eqn:fluxfit}
    \rm F = A\ \frac{(t/t_0)^p}{1 + (t/t_0)^q} + B\ exp(-t/111.26).
\end{equation}

The parameters A, B, $t_0$, p and q in the above function were determined for each SN by minimizing the $\chi^2$ of the fit in the sensitive interval of about 50 days around the time of inflection. The functional fit for 5 type II SNe in our sample along with the parameters $S$ and $t_i$ is shown in Fig.~\ref{fig:steepness}. The best fit for the remaining 25 type II SNe are displayed in Figs.~\ref{fig:miscsteepness1},~\ref{fig:miscsteepness2},~\ref{fig:miscsteepness3},~\ref{fig:miscsteepness4} and~\ref{fig:miscsteepness5}. The derived values of $S$ and $t_i$ for each of the 39 type II SNe used to reconstruct the relation is mentioned along with their respective $\rm ^{56}Ni$ mass from the literature in Table~\ref{tab:lc_steepness}. The best fit to the net sample of type II SNe is shown in Fig.~\ref{fig:nickelsteepness} and is described by the relation:

\begin{equation}
\label{eqn:nickelsteepness}
    \rm log\ M(^{56}Ni) = -(3.5024\pm0.0960) \times S - 1.0167\pm0.0034.
\end{equation}

The $\rm ^{56}Ni$ mass estimated from the steepness relation (\ref{eqn:nickelsteepness}) is also mentioned in Table~\ref{tab:lc_steepness}. The measure of correlation between the two variables (log $\rm M_{Ni})$, $S$) was measured using the Pearson correlation coefficient and the Spearman correlation coefficient which yielded an r-value of --0.8806 and --0.8343, respectively. The chance probability of finding these correlation (p-value) was found to be $\rm 1.47 \times 10^{-11}\ \%$ and $\rm 4.19 \times 10^{-9}\ \%$, respectively.

\subsection{Estimating Nickel mass for \sniip}
\label{sec:calcnickel}

\begin{figure}
\centering
\resizebox{\hsize}{!}{\includegraphics{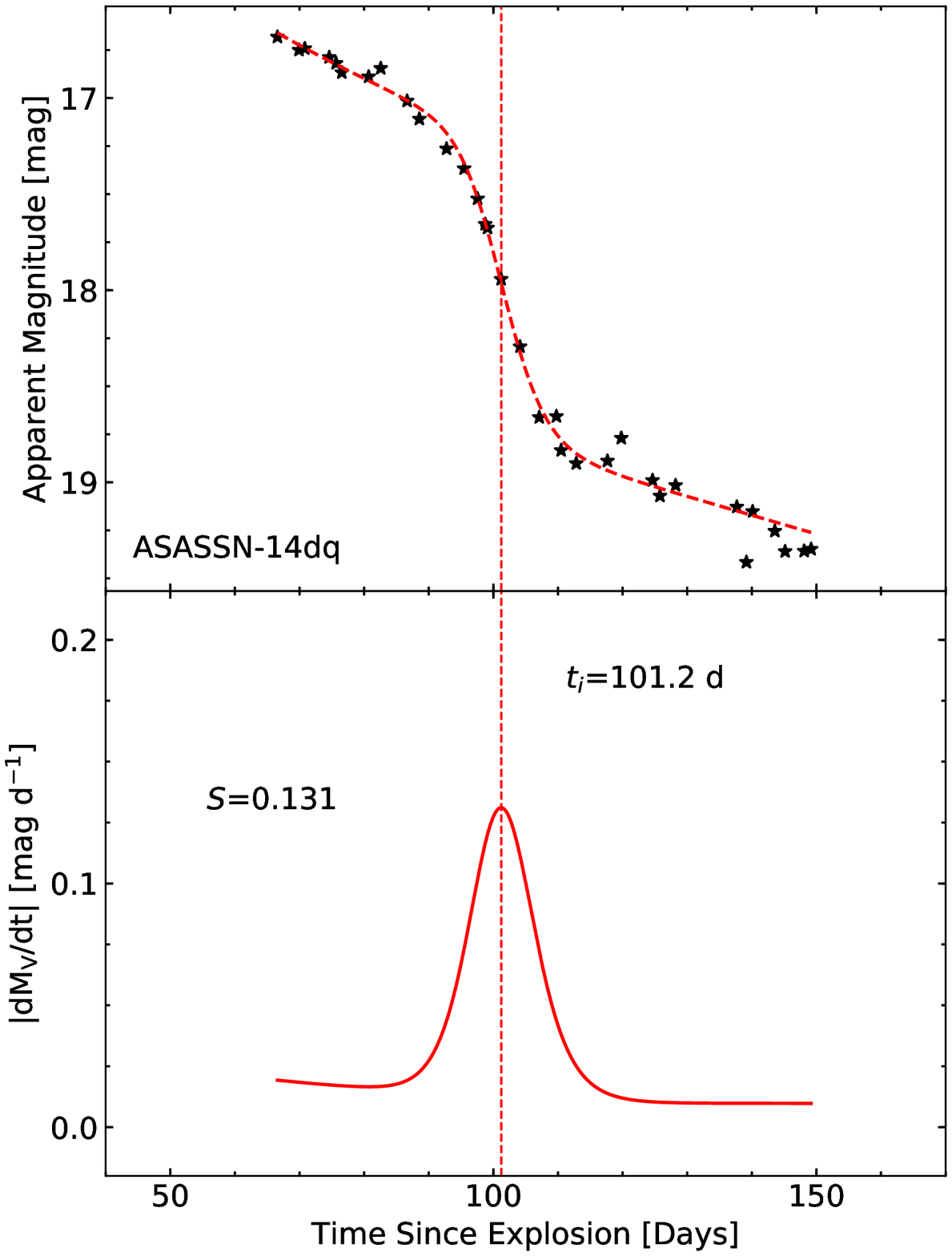}}
\caption{Determination of Steepness parameter and point of inflection for \sniip\ from its $V$-band light curve. The plot description is same as in Fig.~\ref{fig:steepness}.}
\label{fig:14dqsteepness}
\end{figure}

The mass of $\rm ^{56}Ni$ synthesized at the time of the explosion can be independently estimated from the tail bolometric luminosity ($L_t$), as described by \citet{2003hamuy} through the following relation:

\begin{equation}
    \rm M_{Ni} = 7.866 \times 10^{-44} \times L_t \times exp \bigg[\frac{(t_t - t_0)(1 + z) - 6.1}{111.26} \bigg] M_\odot,
\end{equation}

where 6.1 d is the half-life of $\rm ^{56}Ni$, 111.26 d is the e-folding time of $\rm ^{56}Co$ decay and $t_0$ is the explosion time. The tail luminosity, $\rm L_t\,=\,8.33\pm1.15\times 10 ^{40} erg\ s^{-1}$ of \sniip\ determined around $\sim\,$175 d corresponds to a $\rm ^{56}Ni$ mass of $\rm 0.030\pm0.004\ M_{\odot}$.

More massive progenitors of SNe produce a more energetic explosion and SNe with greater explosion energies produce more $\rm ^{56}Ni$ \citep{2003hamuy}. Hence, $\rm ^{56}Ni$ mass can also be computed by comparing the tail bolometric luminosity of the SN with that of SN 1987A, assuming that the $\rm \gamma-ray$ deposition is the same for both SNe. The $\rm ^{56}Ni$ mass for SN 1987A was determined quite accurately by \citet{1998turatto} to be 0.075$\,\pm\,$0.005 $\rm M_{\odot}$. After comparing the tail bolometric luminosity of \sniip\ around $\sim\,$175 d with that of SN 1987A (which is $\rm 2.00\pm0.04\times 10^{41}\ erg\ s^{-1}$), the value of $\rm ^{56}Ni$ mass was found to be $\rm 0.031\pm 0.004\ M_{\odot}$.

The H$\alpha$ luminosity of the SN during the nebular phase can also be used to estimate the $\rm ^{56}Ni$ mass by comparing it to the H$\alpha$ luminosity of SN 1987A at a similar epoch assuming that mass, energy and mixing conditions do not differ strongly \citep{2003elmhamdi}. The H$\alpha$ luminosity of $\rm 3.94\times10^{39}\ erg\ s^{-1}$, derived from the spectrum obtained $\sim\,$176 days from the explosion yields a $\rm ^{56}Ni$ mass of $\rm 0.021\pm0.003\ M_{\odot}$. The mean $\rm ^{56}Ni$ mass estimated using the above methods is $\rm 0.027\pm0.005\ M_{\odot}$.

As discussed earlier in Section~\ref{sec:nickelsteepness}, $\rm ^{56}Ni$ mass can be estimated from the steepness parameter through Eqn.~\ref{eqn:nickelsteepness}. The steepness parameter for \sniip\ is determined in Fig.~\ref{fig:14dqsteepness} and is 0.131$\,\pm\,$0.010 \magday. This yields an $\rm ^{56}Ni$ mass of $\rm 0.033\pm0.003\ M_{\odot}$ using Eqn.~\ref{eqn:nickelsteepness}. This value is similar to the mean value obtained above using other techniques suggesting that \sniip\ obeys the correlation. The average estimated $\rm ^{56}Ni$ mass for \sniip\ using all the above methods is $\rm 0.029\pm0.005\ M_{\odot}$.

%-----------------------------------------------------------------------------%
\section{Light Curve Modelling}
\label{sec:lcmodel}
%-----------------------------------------------------------------------------%

In order to further analyze \sniip, its outburst properties should be determined. The SN outburst is basically characterized by three parameters: the mass of the ejecta ($\rm M_{ej}$), the radius of the star prior to the explosion (R) and the energy of the explosion (E). To determine the above parameters, \citet{2014nagy} proposed a semi-analytic light curve model which assumes a spherically symmetric, homologously expanding SN ejecta. This model is based on the formulation of \citet{1980arnett}, which was further refined by \citet{1989arnett}. For our work, we used the two-component model proposed by \citet{2016nagy} \citep[an extension to][]{2014nagy}, which combines a compact, dense inner core with a low-mass, extended outer envelope \citep[a configuration originally proposed by][to resemble a yellow/red supergiant]{2012bersten}.

The best-fit to the model was obtained by minimizing the $\chi^2$ of the model light curve fit to the observed bolometric light curve. We tried two different values of recombination temperature and two different opacities for the outer envelope (H-rich shell, H-He shell) to see how it affects the values of the parameters obtained due to degeneracies present in the parameter space. The parameters obtained didn't change significantly for different configurations and can be seen in Table~\ref{tab:fitnagy}. Due to the availability of parameters associated with the core in the literature of type II-P SNe, we use only those parameters for discussion. We obtained a radius of $\rm \sim3.6\times10^{13}$ ($\rm \sim500\ R_{\odot}$), an ejecta mass of $\rm \sim10\ M_{\odot}$ and a total energy of $\rm \sim\,1.8\times 10^{51}$ ergs. These values are quite similar to the values obtained for an RSG progenitor of SN 2013ej by \citet{2016nagy} and \citet{2015boseej}. This indicates similarity in the progenitors of \sniip\ and SN 2013ej. The mass of $\rm ^{56}$Ni estimated from the two-component model using the nebular phase in the light curve is 0.026 M$_{\odot}$ and is very similar to our $\rm ^{56}Ni$ mass estimate in Section~\ref{sec:calcnickel}.

\begin{table*}
\centering
\caption{Parameters estimated from two-component semi-analytic modelling of the bolometric light curve of \sniip\ proposed by \citet{2016nagy}.} 
\label{tab:fitnagy}                     
\begin{tabular}{l c c c}          
\hline
\hline                      
Parameter                                   &  Core (He-rich)               & Shell (mixed H-He)                & Shell (pure H) \\ 
                                            & ($\kappa$\,=\,0.2 cm$^2$/g)   &  ($\kappa$\,=\,0.3 cm$^2$/g)      & ($\kappa$\,=\,0.4 cm$^2$/g) \\
\noalign{\smallskip} \hline \noalign{\smallskip}                  
$R_{0}$ ($\times 10^{12}$cm), Initial radius of the ejecta  & 38            & 72                                & 74    \\
$T_{rec}$ (K), Recombination temperature    & 5500                          & --                                & --    \\
$M_{ej}$ (M$_{\odot}$), Ejecta mass         & 10.3                          & 0.68                              & 0.59  \\
$M_{Ni}$ (M$_{\odot}$), Nickel mass         & 0.026                         & --                                & --    \\
$E_{Th}$ (foe), Thermal energy              & 0.38                          & 0.09                              & 0.09  \\
$E_{kin}$ (foe), Kinetic energy             & 1.65                          & 1.54                              & 1.56  \\
\noalign{\smallskip} \hline \noalign{\smallskip}
$R_{0}$ ($\times 10^{12}$cm), Initial radius of the ejecta  & 36            & 76                                & 72    \\
$T_{rec}$ (K), Recombination temperature    & 7000                          & --                                & --    \\
$M_{ej}$ (M$_{\odot}$), Ejecta mass         & 10.4                          & 0.66                              & 0.70  \\
$M_{Ni}$ (M$_{\odot}$), Nickel mass         & 0.026                         & --                                & --    \\
$E_{Th}$ (foe), Thermal energy              & 0.37                          & 0.10                              & 0.09  \\
$E_{kin}$ (foe), Kinetic energy             & 1.24                          & 1.46                              & 1.30  \\
\noalign{\smallskip} \hline
\end{tabular}
\smallskip
\end{table*}

\begin{figure*}
\centering
\resizebox{\hsize}{!}{\includegraphics{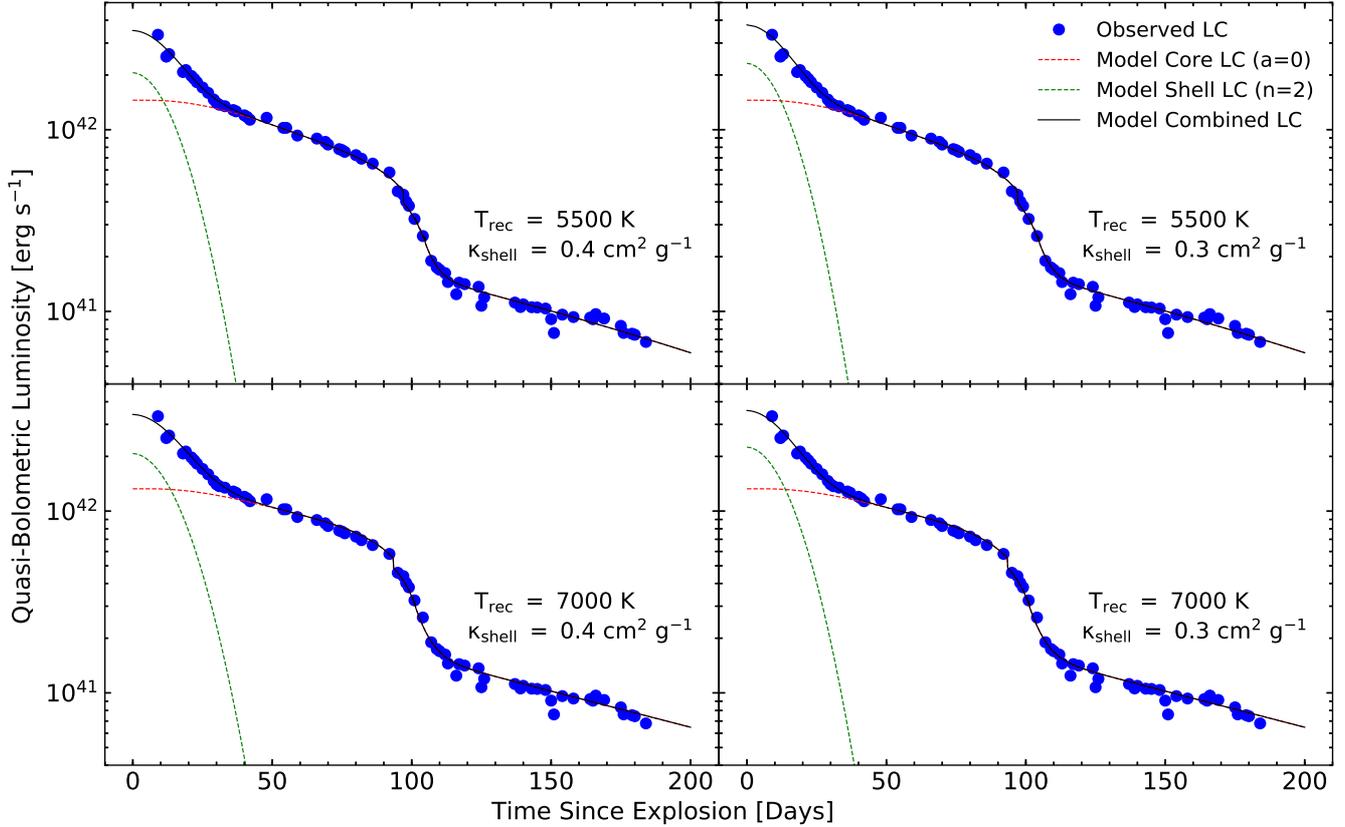}}
\caption{Fitting of observed bolometric light curve of \sniip\ with a two-component model \citep{2016nagy}. The model light curve comprises of light curves originating from a compact core and an extended envelope. The plots differ in the choice of opacity chosen for the extended envelope (shell): H-rich ($\rm \kappa\,=\,0.4\ cm^{2}\ g^{-1}$) and mixed H-He ($\rm \kappa\,=\,0.3\ cm^{2}\ g^{-1}$) and the choice of recombination temperature ($\rm T_{rec}$) for the core.}
\label{fig:fitnagy}
\end{figure*}

%------------------------------------------------------------------------------%
\section{\sniip\ among type II SNe}
\label{sec:disc}
%------------------------------------------------------------------------------%

\begin{table*}
\centering
\caption{Parameters of type II SNe referred in this paper \citep[in addition to the previous works of][]{2003hamuy, 2014spiro, 2015valenti}.}
\setlength{\tabcolsep}{5pt}
\label{tab:snpar}                     
\begin{tabular}{l c c c c c c}          
\hline \hline                      
SN           &  Explosion Epoch &   Distance Modulus, $\mu$     &   $\rm M^V_{50}$          &   $\rm M_{Ni}$                    &   $\rm V^{50}_{FeII}$     &   Reference   \\
(Name)       &  (JD)            &   (mag)                       &   (mag)                   &   ($\rm M_{\odot}$)               &   (\kms)                  &               \\
\noalign{\smallskip} \hline \noalign{\smallskip}
2003gd       &  2452717.00      &   29.84$\,\pm\,$0.36          &  -15.92$\,\pm\,$0.49      &   0.016$^{+0.010}_{-0.006}$       &   3694$\,\pm\,$981        &   1           \\
2009ib       &  2455041.30      &   31.48$\,\pm\,$0.31          &  -16.28$\,\pm\,$0.31      &   0.046$\,\pm\,$0.015             &   3247$\,\pm\,$200        &   4           \\
2012ec       &  2456143.00      &   31.19$\,\pm\,$0.13          &  -16.54$\,\pm\,$0.14      &   0.040$\,\pm\,$0.015             &   3700$\,\pm\,$100        &   2           \\
2013ab       &  2456340.00      &   31.92$\,\pm\,$0.10          &  -16.70$\,\pm\,$0.10      &   0.064$\,\pm\,$0.006             &   4400$\,\pm\,$400        &   3           \\
ASASSN-14ha  &  2456910.50      &   29.53$\,\pm\,$0.50          &  -14.40$\,\pm\,$0.50      &   0.0014$\,\pm\,$0.0002           &       --                  &   5           \\       
2016X        &  2457405.92      &   30.91$\,\pm\,$0.43          &  -16.20$\,\pm\,$0.43      &   0.034$\,\pm\,$0.006             &   4500$\,\pm\,$200        &   6           \\   
2016bkv	     &  2457467.50      &   30.79$\,\pm\,$0.04          &  -14.74$\,\pm\,$0.06      &   0.022$\,\pm\,$0.001             &   1300$\,\pm\,$150        &   8           \\
2017eaw	     &  2457884.00      &   28.74$\,\pm\,$0.11$^{a}$    &  -16.44$\,\pm\,$0.12      &   0.050                           &       --                  &   7           \\
\noalign{\smallskip} \hline \noalign{\smallskip} 
\end{tabular}
\newline
$\rm ^{a}$The distance to the galaxy NGC 6946 hosting the SN 2017eaw was taken from \citet{2006sahu}. \\
(1) \citet{2005hendry}
(2) \citet{2015barbarino};
(3) \citet{2015boseab};
(4) \citet{2015takats};
(5) \citet{2016valenti};
(6) \citet{2018huang};
(7) \citet{2018tsvetkov};
(8) \citet{2018hosseinzadeh}.
\end{table*}

\begin{figure}
\centering
\resizebox{\hsize}{!}{\includegraphics{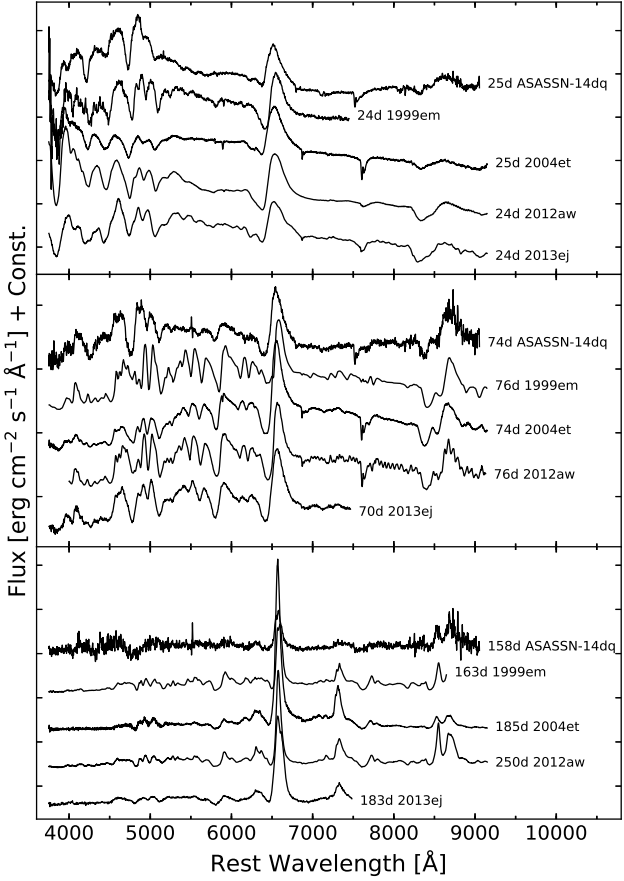}}
\caption{Comparison of spectra of \sniip\ with other well studied type II SNe at epochs $\sim\,$24 d, $\sim\,$74 d and $\sim\,$160 d. All the spectra have been corrected for total extinction and redshift. References for our sample of comparison is listed in Section~\ref{sec:applc}.}
\label{fig:speccomp}
\end{figure}

\begin{figure}
\centering
\resizebox{\hsize}{!}{\includegraphics{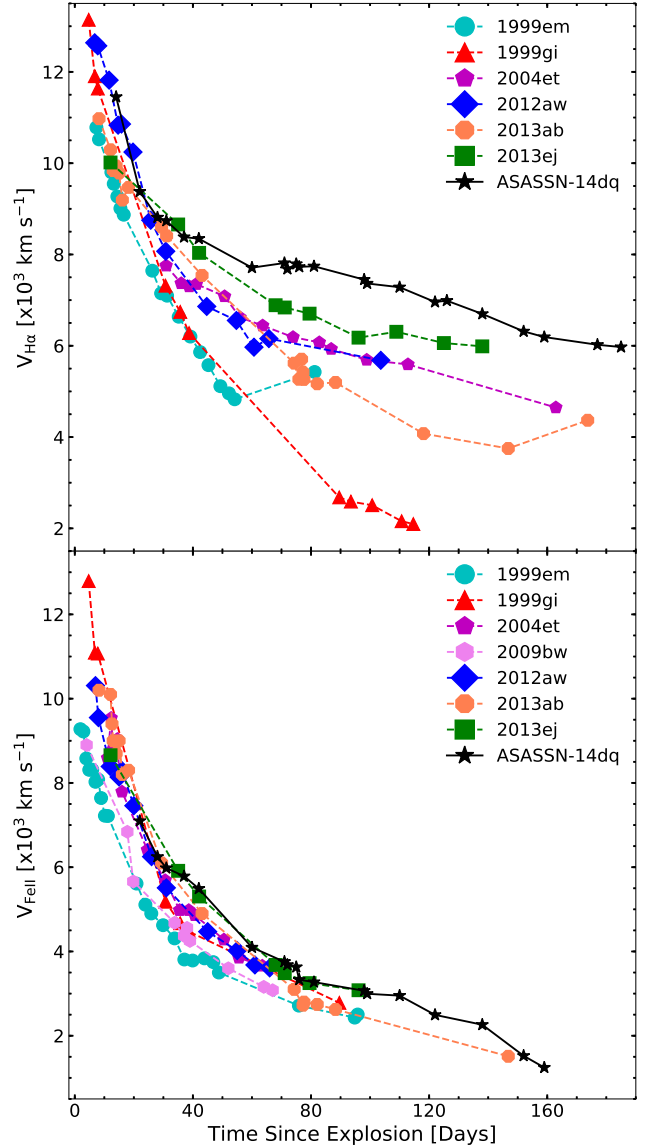}}
\caption{Velocity evolution of \sniip\ compared with other well-studied type II SNe. The velocities are computed from the FWHM of the absorption troughs of the P-Cygni profiles. The \textit{top} panel consists of velocities inferred from H$\alpha$ whereas the \textit{bottom} panel from the Fe\,II $\lambda$5169 \AA\ feature. The data has been taken from the references listed in Section~\ref{sec:applc} and \citet{2014bose}.}
\label{fig:compphotvel}
\end{figure}

\begin{figure*}
\centering
\resizebox{\hsize}{!}{\includegraphics{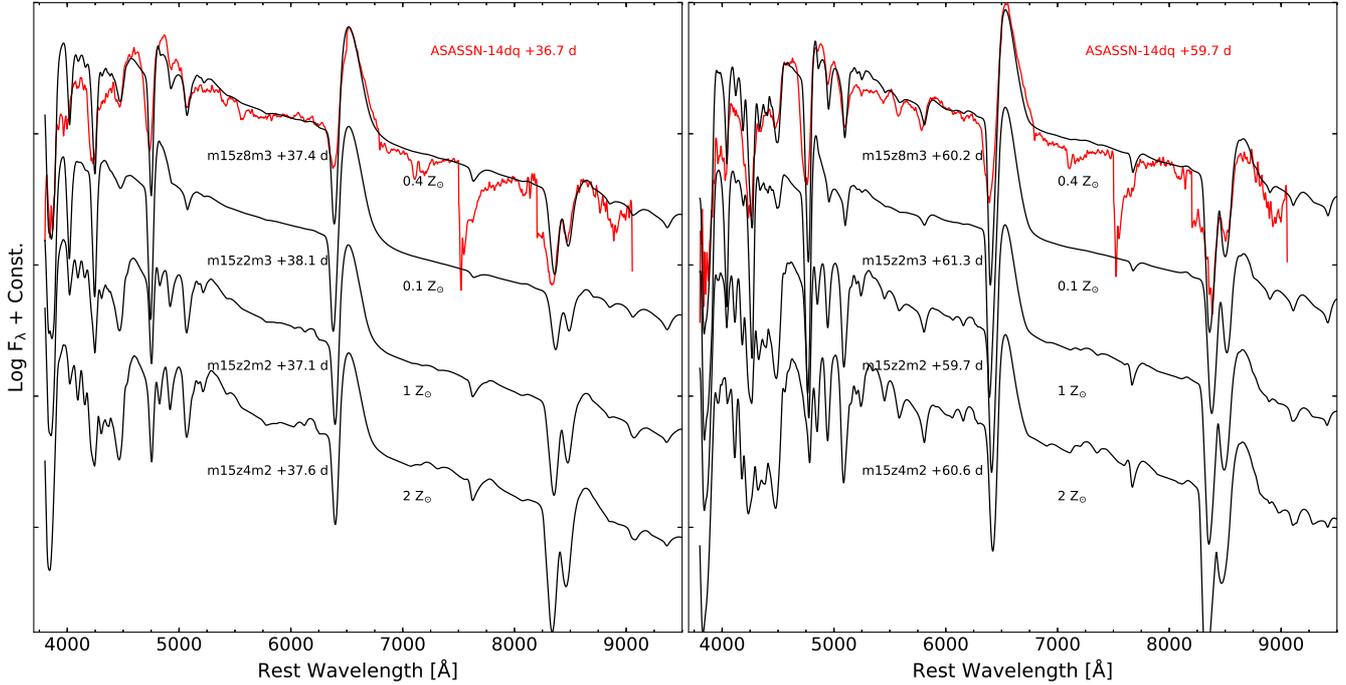}}
\caption{Comparison of \sniip\ spectra with the models of \citet{2013bdessart} of different metallicities. The epochs of comparison are $\sim\,$37 and $\sim\,$60 days from the date of explosion. The model with 0.4 $\rm Z_{\odot}$ (m15z8m3) matches best with the spectrum of \sniip\ at both the epochs of comparison.}
\label{fig:compmodspec}
\end{figure*}

\begin{figure}
\centering
\resizebox{\hsize}{!}{\includegraphics{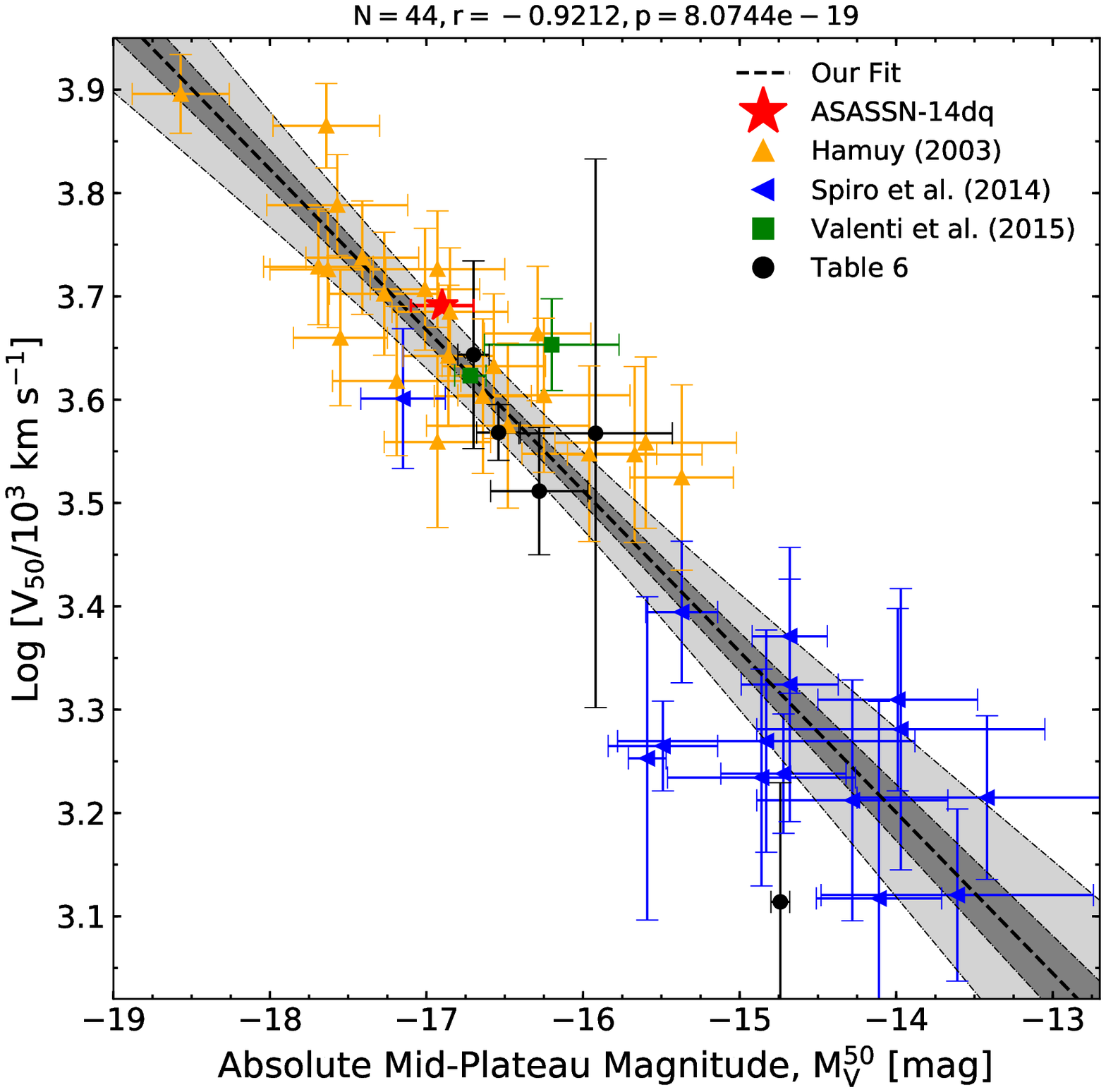}}
\caption{Plot of photospheric velocity ($\rm Log\ V_{50}$) vs mid-plateau $V$-band absolute magnitude ($\rm M_{V}^{50}$) for type II SNe. Data has been taken from \citet{2003hamuy}, \citet{2014spiro}, \citet{2015valenti} and Table~\ref{tab:snpar}. The black-dotted line displays the fit for the type II SNe plotted. The photospheric velocity was computed from the Fe\,II 5018 \AA\ absorption feature. References are the same as in Fig.~\ref{fig:ni}. The 1$\sigma$ and 3$\sigma$ confidence intervals of the fit are shaded in $dark-grey$ and $light-grey$, respectively. The Pearson correlation coefficient along with the chance probability of finding a correlation is mentioned at the top of the figure.}
\label{fig:mv}
\end{figure}

\begin{figure}
\centering
\resizebox{\hsize}{!}{\includegraphics{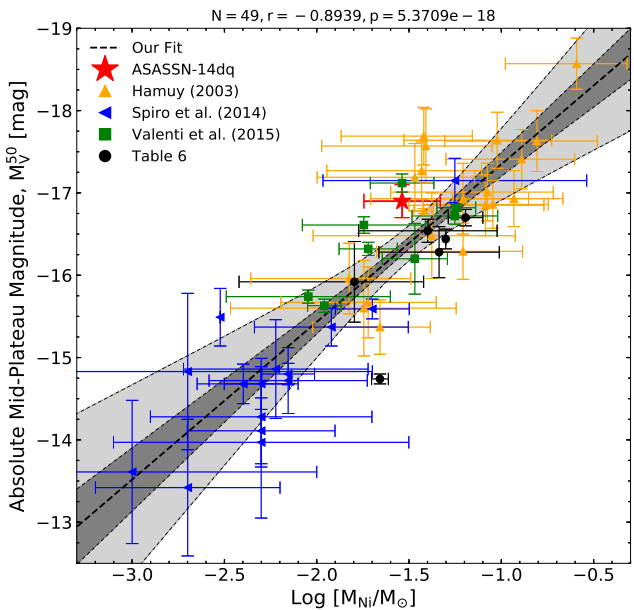}}
\caption{Plot of mid-plateau $V$-band absolute magnitude ($\rm M_{V}^{50}$) vs nickel mass ($\rm Log\ M_{Ni}$) for type II SNe. References and plot description is the same as in Fig.~\ref{fig:mv}.}
\label{fig:ni}
\end{figure}

\begin{figure}
\centering
\resizebox{\hsize}{!}{\includegraphics{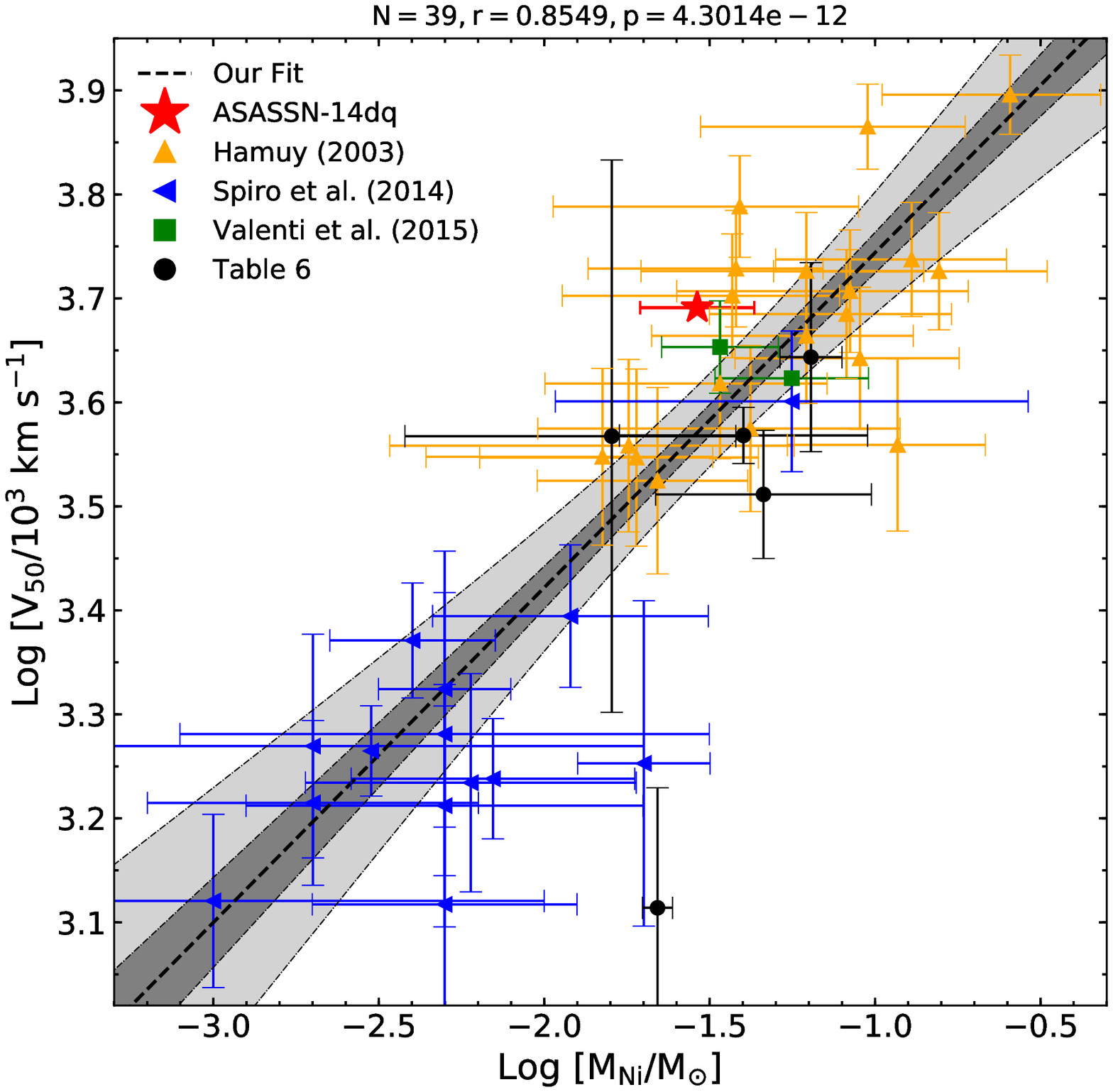}}
\caption{Plot of photospheric velocity ($\rm Log\ V_{50}$) vs nickel mass ($\rm Log\ M_{Ni}$) for type II SNe. The photospheric velocity was computed from Fe\,II 5018 \AA\ absorption feature. References and plot description is the same as in Fig.~\ref{fig:mv}.}
\label{fig:v50}
\end{figure}

The light curves of \sniip\ display a well-sampled transitional phase from the plateau to the radioactive-tail. The steepness parameter for \sniip, 0.131$\,\pm\,$0.010 \magday\ (estimated earlier in Section~\ref{sec:calcnickel}) is similar to the value inferred for type II SNe like SN 1999em (0.118) and SN 1999gi (0.115). If we compare \sniip\ with luminous type II SNe, the steepness is less than that of SN 2013ej (0.181) and SN 2013by (0.174), but is more than in the case of SN 2004et(0.076), SN 1970G (0.085) and SN 1992H (0.048). This signifies the diversity in type II SNe. In Section~\ref{sec:nickelsteepness}, we found that the correlation between the steepness parameter $S$ and $\rm ^{56}Ni$ mass was strengthened (low p-value) after the inclusion of more type II SNe to the original sample used in \citet{2003elmhamdi}. Also, the correlation strengthens the fact that increase in the $\rm ^{56}Ni$ mass in the SN ejecta favours radiative diffusion at the end of the plateau and causes a less steep transition to the radioactive-tail \citep{2003elmhamdi}. However, we do expect deviations from this correlation due to differences in the degree of $\rm ^{56}Ni$ mixing. The effect of higher degree of $\rm ^{56}Ni$ mixing can cause decreased steepness in the transition phase because of an increase in radiative diffusion \citep[see Fig. 12 in][]{1994eastman}.

In Fig.~\ref{fig:speccomp}, the spectra of \sniip\ are compared with the normal type II-P SN 1999em \citep{2002bleonard}, luminous type II-P SN 2004et \citep{2006sahu}, the intermediate-luminosity type II-P SN 2012aw \citep{2013bose} and the luminous type II-P/L SN 2013ej \citep{2015boseej} at three different epochs ($\sim\,$24 d, $\sim\,$74 d and $\sim\,$160 d). Even though the spectra of \sniip\ match quite well with other type II SNe at all the three epochs, it appears relatively featureless and displays weaker metal lines in comparison. This might be due to the low (sub-solar) oxygen abundance of the host galaxy estimated in Section~\ref{sec:prophost}. To further investigate the possibility of low host metallicity, we compare the spectra of \sniip\ with the models of D13 generated from the explosion of a 15 $\rm M_{\odot}$ main-sequence star with MESA\footnote{\url{http://mesa.sourceforge.net/}} \citep{2011mesa, 2013mesa} at 0.1, 0.4, 1 and 2 $\rm Z_{\odot}$ metallicities. In the photospheric phase, D14 found that the observed intermediate-mass elements along with the Fe-group elements reflect the original composition of the progenitor star and are weakly affected by nuclear burning in the core.

In Fig.~\ref{fig:compmodspec}, the spectra of \sniip\ (at two epochs) is compared with the models of D13. The spectra shows resemblance to the model spectra with 0.4 $\rm Z_{\odot}$ metallicity (m15z8m3) in terms of pseudo-Equivalent Width (pEW) of the metal lines (Ca\,II, Fe\,II etc.). It is important to note here that the models of D13 are not tailored for the progenitor of ASASSN-14dq. The Balmer absorption is weaker in the spectra of \sniip\ than in the models of D13 and hints towards a slower evolution of \sniip. This might result from the residual thermal energy of an earlier interaction of the outer ejecta with the CSM \citep{2016polshaw}. Although type II-P SNe typically show deeper P-Cygni absorption profiles \citep{1996schlegel, 2014gutierrez}, the absorption troughs of the H$\alpha$ P-Cygni profile in the spectra of \sniip\ are shallower and are comparable with the type II-P/L SN 2013ej. This indicates a hydrogen-poor envelope above the photosphere leading to a prominent emission feature but less significant absorption feature and could possibly be a result of low metallicity of the host \citep{2018gutierrez}. Type II SNe in large late-type spirals are on an average brighter (an average peak magnitude of --16.28$\,\pm\,\,$0.35, $\sigma$\,=\,1.52 mag) than those in the smaller ones \citep{2011li}. This is also the case with \sniip, which is brighter than an average type II-P SN. However, \citet{2018gutierrez} do not find any differences in the absolute magnitude of type II SNe in different galaxies.

In Fig.~\ref{fig:compphotvel}, the velocity evolution of \sniip\ inferred from two different absorption features (H$\alpha$, Fe\,II $\lambda$5169 \AA) is compared with other well-studied type II SNe. The rate of decline of H$\alpha$ line velocity is similar to other type II SNe in the early phase but is consistently slower after the mid-plateau phase ($\sim\,$45 d). The power law ($v\,=\,a*t^{-b}$) fit to the H$\alpha$ and the H$\beta$ velocity evolution returned exponents 0.18 and 0.32, respectively, which are lesser compared to the values (0.41 for H$\alpha$ and 0.53 for H$\beta$) obtained by \citet{2014afaran} for type II-P SNe. This suggests a hydrogen-poor envelope in \sniip\ and a possible interaction of the ejecta with the CSM. The evolution of line velocities inferred from the Fe\,II triplet for \sniip\ matches very well with the other type II SNe throughout the plateau phase and was confirmed by the power law fit to the evolution of Fe\,II line feature, which declined with an exponent of 0.59 \citep[compared to 0.58 in][for type II-P SNe]{2014afaran}. The less scatter seen among the velocities inferred from Fe\,II absorption feature is consistent with the inference of \citet{2005bdessart}, which emphasized only $\sim\,$5-10~\% errors on photospheric velocity determination from Fe\,II absorption features, whereas the scatter in the H$\alpha$ line velocities increases as the SN ages across the plateau phase. \citet{2014bfaran} have shown that type II-P SNe display fast evolving (fast-declining) H$\beta$ velocity evolution due to their thick hydrogen envelope and drop by a factor of $\sim\,$3 from the peak to the mid-plateau phase compared to type II-L SNe that drop $\leq$\,50~\%. \sniip\ shows a relatively slow evolution (as seen in Fig.~\ref{fig:photvel}), although it is not as slow as a type II-L SN.

To compare the observed parameters of \sniip\ with type II SNe, we derive a sample from \citet{2003hamuy}, \citet{2014spiro} and \citet{2015valenti} along with 8 additional events mentioned in Table~\ref{tab:snpar}. In Fig.~\ref{fig:mv}, the mid-plateau $V$-band absolute magnitude ($\rm M_{V}^{50}$) is plotted with the mid-plateau photospheric velocity ($\rm Log\ V_{50}$) for our sample of type II SNe. We find that \sniip\ lies within the 3$\sigma$ confidence interval of the fit. In Figs.~\ref{fig:ni} and \ref{fig:v50}, $\rm ^{56}Ni$ mass synthesized in the explosion is plotted against $\rm M_{V}^{50}$ and $\rm Log\ V_{50}$ of our sample, respectively. \sniip\ lies outside the 3$\sigma$ confidence interval of both the fits associated with the $\rm ^{56}Ni$ mass, indicating that \sniip\ has a lower yield of $\rm ^{56}Ni$ for a type II SN. SN 1986I, SN 1969L and 1999cr have similar mid-plateau magnitude but have significantly ($\sim\,$3-4 times) higher $\rm ^{56}Ni$ mass synthesized in the explosion. SN 2003hn is the only type II SN that comes close to \sniip\ in terms of $\rm^{56}Ni$ mass synthesized for a similar plateau brightness apart from the transitional type II-P/L SN 2013ej which is slightly fainter ($\sim\,$0.3 mag) in the plateau phase when compared to \sniip.

Following the classification of \citet{2014bfaran}, where type II-P SNe decline with a rate of s$\rm 50_V\,<\,0.5$, \sniip\ (with an s$\rm 50_V\,\sim\,$0.80) would be classified as a type II-L SN. However, if we use the classification criteria used by \citet{2015valenti} for a type II-P SN as having s$\rm 50_V\,\leq\,$1.0, \sniip\ will be classified as a type II-P SN. Nonetheless, the presence of a drop in magnitude at the end of the hydrogen recombination phase is usually considered the defining feature of a type II-P SN \citep{2015valenti} and can be seen in \sniip.

In a large sample study of type II SNe, \citet{2014anderson} and \citet{2015sanders} found no clear photometric gap that distinguishes the type II-P SNe from the type II-L SNe. The plateau durations of type II-P SNe are typically $\sim\,$100 d but get shorter for more luminous type II SNe \citep{2013poznanski} and display high-velocity H$\alpha$ absorption in the spectra suggesting an interaction with the CSM. The high decline rate of \sniip\ during the plateau phase, inferred earlier in Section~\ref{sec:lightcurve}, matches with the short plateau of \sniip\ and is consistent with the correlation between the plateau length and the plateau decline rate for type II-P SNe inferred by \citet{2014anderson}. Type II SNe which decline rapidly at early epochs also generally decline rapidly during the plateau and in the radioactive phase \citep{2014anderson}. The same has been observed for \sniip. The late phase light curve in type II SNe is powered by the radioactive decay of $\rm ^{56}Co$ to $\rm ^{56}Fe$, which in the case of complete $\gamma$-ray trapping and $e^{+}$ trapping has a characteristic decline rate of 0.98 \maghundred. In the nebular phase, \sniip\ shows a decline rate ($s3$) comparable to the decay rate of $\rm ^{56}Co$ and hence indicates complete $\gamma$-ray trapping.

The low $\rm ^{56}Ni$ mass inferred for \sniip\ is in contrast with the brighter maximum luminosity, indicating an alternate source of energy, such as the interaction of the ejecta with the CSM. The observed asymmetry in the nebular H$\alpha$ emission also indicates a possible CSM interaction. Asymmetric line profiles have also been observed in other type II-P SNe such as SN 1999em \citep{2002aleonard}, SN 2004dj \citep{2005chugai} and in SN 2013ej \citep{2015boseej}. \citet{2005chugai} suggest that the asymmetry is due to an asymmetric ejection of $\rm ^{56}Ni$ caused by the interaction of the ejecta with a geometrically asymmetric CSM. An asymmetric bipolar $\rm ^{56}Ni$ ejecta is not an exception for type II-P SNe and can result from a bipolar explosion in CCSNe due to a mildly rotating RSG core \citep{2005chugai}. The fast decline rate of \sniip\ during the plateau phase may also arise from asymmetric CSM interaction as inferred for PTF11iqb \citep{2015smith}.

The photometric and spectroscopic parameters of \sniip\ suggests that it is a luminous, fast declining type II-P SN. \citet{2014anderson} showed that progenitors of fast-declining type II SNe retain a smaller amount of hydrogen in the envelope than the ones with slow decline rates. Type II SNe with faster decline rates display faster release of energy and brighter maximum luminosity due to low ionization and ejecta expansion \citep{1994patat}.

%------------------------------------------------------------------------------%
\section{Summary}
\label{sec:sum}
%------------------------------------------------------------------------------%

We have presented 34 epochs of broadband ($UBVRI$) optical photometry along with 23 epochs of low-resolution spectroscopic observations of \sniip\ during the first 195 days of its evolution. The apparent magnitude light curves and the spectra of \sniip\ suggest that it is a type II-P SN exhibiting a plateau which lasts about $\sim\,$90 days. The peak $V$-band absolute magnitude (i.e 17.7$\,\pm\,$0.2) and the peak bolometric luminosity (i.e $\rm \sim2.6\times10^{42} erg\ s^{-1}$) of \sniip\ classifies it as a luminous, fast-declining type II-P SN. We inferred negligible reddening for \sniip\ from the dwarf host galaxy \hostglx. We adopted a total minimal reddening of $\rm E(B-V)$\,=\,0.06 mag for \sniip\ owing completely to the ISM in the Milky Way.

We inferred a distance of 44.8$\,\pm\,$3.1 Mpc to \sniip. We also estimated an oxygen abundance of 12 + log(O/H)\,=\,8.40$\,\pm\,$0.18 of the host galaxy \hostglx. A sub-solar oxygen abundance of the host galaxy (and hence the progenitor) was reflected in our relatively featureless (and low pEW) spectra of \sniip. The spectra also matched closely with the D13 model spectra of sub-solar metallicity (0.4 $\rm Z_{\odot}$). This inference supports the correlation between progenitor metallicity with the metal line pEW in the photospheric phase of the type II SNe.

We rebuilt the empirical relation between the steepness parameter and the $\rm ^{56}Ni$ mass synthesized in the explosion \citep{2003elmhamdi} using well-sampled $V$-band light curves of type II SNe from the literature. The larger sample in our study (including low-luminosity type II SNe) strengthened the correlation between the two parameters and indicated towards a slightly shallower fit (Eqn.~\ref{eqn:nickelsteepness}) than that of \citet{2003elmhamdi}. The decline rate of \sniip\ in the late stages of the plateau phase (s2\,=\,1.18 \maghundred) is close to the mean decline rate (1.27 \maghundred) inferred for all of type II SNe in the sample of \citet{2014anderson}. The $\rm ^{56}Ni$ mass synthesized during the explosion is estimated to be 0.029$\,\pm\,$0.005 $\rm M_\odot$ and is slightly ($\sim\,$50\%) lower for a luminous type II SN. We obtained the explosion parameters of \sniip\ through analytical modelling of its bolometric light curve. We obtained an explosion energy, $\rm E_{exp} \sim\,1.8\times10^{51}\ erg$, mass of the ejecta, $\rm M_{ej}\,=\,\sim\,10\ M_{\odot}$. and a radius, $\rm R\,=\,\sim\,500\ R_{\odot}$.

In the spectroscopic evolution of \sniip, prominent P-Cygni profiles of H$\alpha$, H$\beta$, Fe\,II 5169, 5018, 4924 \AA, He\,I 5876 \AA, Ca\,II 8498, 9042, 8662 etc. were observed as in a type II-P SN. The H$\alpha$ P-Cygni profile in the early phase spectra of \sniip\ displays a shallower absorption feature compared to other type II-P SNe, indicating a hydrogen-poor envelope. An HV component is seen in the early phase spectra with a velocity of $\sim\,$17000 \kms\ in the spectrum of $\sim\,$19 d, which completely disappears after $\sim\,$42 d. The velocity evolution of Fe\,II feature is quite similar to that of a type II-P SN, although the velocities inferred from Balmer absorption features decline slowly for a type II-P SN. The asymmetry in the H$\alpha$ profile, the low $\rm ^{56}Ni$ mass inferred, the bright plateau luminosity and the fast decline during the plateau phase points towards the interaction of the ejecta with an asymmetric CSM. All the above inferences indicate a hydrogen-poor envelope in the progenitor of \sniip, which may have lost most of its hydrogen during pre-supernova evolution. This makes it a transitional event between the type II-P and II-L SNe.

%------------------------------------------------------------------------------%
\section*{Acknowledgements}
%------------------------------------------------------------------------------%

We thank the anonymous referee for thorough reading of the manuscript and for his/her helpful suggestions which helped in improving the content and readability of the manuscript. We thank Subhash Bose who provided us with the photospheric velocity data for SN 2013ej and SN 2013ab. We thank Chayan Mondal and Mousumi Das of the Indian Institute of Astrophysics for their helpful suggestions on morphology and metallicity of galaxies. We thank the staff at IAO, Hanle and CREST, Hoskote who helped with the observations from the 2m HCT. The facilities at IAO and CREST are operated by the Indian Institute of Astrophysics, Bengaluru. We also thank all the observers who helped us with the follow-up observations of the SN by providing a part of their observing time. Brajesh Kumar also acknowledges the Science and Engineering Research Board (SERB) under the Department of Science \& Technology, Govt. of India, for financial assistance in the form of National Post-Doctoral Fellowship (Ref. no. PDF/2016/001563). This work made use of the NASA Astrophysics Data System and the NASA/IPAC Extragalactic Database (NED) which is operated by the Jet Propulsion Laboratory, California Institute of Technology. 

This work also made use of the data from the Pan-STARRS1 Survey (PS1). The PS1 public science archive have been made possible through contributions by the Institute for Astronomy, the University of Hawaii, the Pan-STARRS Project Office, the Max-Planck Society and its participating institutes, the Max Planck Institute for Astronomy, Heidelberg and the Max Planck Institute for Extraterrestrial Physics, Garching, The Johns Hopkins University, Durham University, the University of Edinburgh, the Queen's University Belfast, the Harvard-Smithsonian Center for Astrophysics, the Las Cumbres Observatory Global Telescope Network Incorporated, the National Central University of Taiwan, the Space Telescope Science Institute, the National Aeronautics and Space Administration under Grant No. NNX08AR22G issued through the Planetary Science Division of the NASA Science Mission Directorate, the National Science Foundation Grant No. AST-1238877, the University of Maryland, Eotvos Lorand University (ELTE), the Los Alamos National Laboratory, and the Gordon and Betty Moore Foundation.
%------------------------------------------------------------------------------%

\label{lastpage}

\bibliographystyle{mnras}
\bibliography{_Reference}

\appendix

\section{Tables}

\begin{table*}
\centering
\setlength{\tabcolsep}{3pt}
\caption{Log of spectroscopic observations of \sniip\ from HCT.}
\label{tab:speclog}
\begin{tabular}{c c c c}
\hline \hline
     Date     &     JD      & Phase$^*$ &        Range           \\
 (yyyy-mm-dd) & (2456700+)  &  (d)      &        (\AA)           \\
 \hline \noalign{\smallskip}
 2014-07-16   &  155.38     &  +13.88   & 5200--9250             \\
 2014-07-21   &  160.31     &  +18.81   & 3500--7800             \\
 2014-07-24   &  163.35     &  +21.85   & 3500--7800             \\
 2014-07-30   &  169.35     &  +27.85   & 3500--7800; 5200--9250 \\
 2014-08-02   &  172.29     &  +30.79   & 3500--7800; 5200--9250 \\
 2014-08-08   &  178.21     &  +36.71   & 3500--7800; 5200--9250 \\
 2014-08-13   &  183.42     &  +41.92   & 3500--7800; 5200--9250 \\
 2014-08-31   &  201.20     &  +59.70   & 3500--7800; 5200--9250 \\
 2014-09-11   &  212.32     &  +70.82   & 3500--7800; 5200--9250 \\
 2014-09-12   &  213.31     &  +71.81   & 3500--7800; 5200--9250 \\
 2014-09-15   &  216.10     &  +74.60   & 3500--7800             \\
 2014-09-16   &  217.16     &  +75.66   & 3500--7800; 5200--9250 \\
 2014-09-21   &  222.20     &  +80.70   & 3500--7800; 5200--9250 \\
 2014-10-08   &  239.06     &  +97.56   & 3500--7800             \\
 2014-10-09   &  240.23     &  +98.73   & 3500--7800; 5200--9250 \\
 2014-10-20   &  251.26     & +109.76   & 3500--7800             \\
 2014-11-01   &  263.02     & +121.52   & 3500--7800; 5200--9250 \\
 2014-11-05   &  267.13     & +125.63   & 3500--7800; 5200--9250 \\
 2014-11-17   &  279.18     & +137.68   & 3500--7800; 5200--9250 \\
 2014-12-01   &  293.09     & +151.59   & 3500--7800; 5200--9250 \\
 2014-12-08   &  300.08     & +158.58   & 3500--7800; 5200--9250 \\
 2014-12-26   &  318.04     & +176.54   & 3500--7800; 5200--9250 \\
 2015-01-03   &  326.04     & +184.54   & 3500--7800             \\
\noalign{\smallskip} \hline
\multicolumn{3}{l}{$^*$\footnotesize{Time since explosion epoch (JD 2456841.50)}.}
\end{tabular}
\end{table*}
\begin{table*}
\centering
\renewcommand{\arraystretch}{1.1}
\caption{$UBVRI$ magnitudes of secondary standards in the SN field.}
\label{tab:secstd}
\vspace{3mm} 
\begin{tabular}{c c c c c c}
\hline \hline
ID &       U            &       B               &       V           &       R               &       I           \\
   &   (mag)            &   (mag)               &   (mag)           &   (mag)               &   (mag)           \\
\hline \noalign{\smallskip}
1  &  16.12$\,\pm\,$0.03 &  16.08$\,\pm\,$0.01 &  15.46$\,\pm\,$0.01 &  15.08$\,\pm\,$0.01 &  14.71$\,\pm\,$0.01 \\
2  &  17.11$\,\pm\,$0.03 &  16.46$\,\pm\,$0.01 &  15.56$\,\pm\,$0.01 &  15.07$\,\pm\,$0.01 &  14.60$\,\pm\,$0.01 \\
3  &  17.37$\,\pm\,$0.03 &  17.27$\,\pm\,$0.01 &  16.60$\,\pm\,$0.01 &  16.20$\,\pm\,$0.02 &  15.79$\,\pm\,$0.02 \\
4  &  16.72$\,\pm\,$0.03 &  16.07$\,\pm\,$0.01 &  15.13$\,\pm\,$0.01 &  14.62$\,\pm\,$0.01 &  14.15$\,\pm\,$0.01 \\
5  &  16.41$\,\pm\,$0.03 &  16.35$\,\pm\,$0.01 &  15.71$\,\pm\,$0.01 &  15.31$\,\pm\,$0.01 &  14.91$\,\pm\,$0.01 \\
6  &  19.37$\,\pm\,$0.04 &  18.04$\,\pm\,$0.01 &  16.56$\,\pm\,$0.01 &  15.59$\,\pm\,$0.01 &  14.65$\,\pm\,$0.02 \\
7  &  15.21$\,\pm\,$0.03 &  14.65$\,\pm\,$0.01 &  13.77$\,\pm\,$0.01 &  13.36$\,\pm\,$0.01 &  12.93$\,\pm\,$0.01 \\
8  &  14.34$\,\pm\,$0.03 &  14.39$\,\pm\,$0.01 &  13.86$\,\pm\,$0.01 &  13.56$\,\pm\,$0.01 &  13.21$\,\pm\,$0.01 \\
9  &  15.62$\,\pm\,$0.03 &  15.34$\,\pm\,$0.01 &  14.55$\,\pm\,$0.01 &  14.10$\,\pm\,$0.01 &  13.64$\,\pm\,$0.01 \\
10 &  15.09$\,\pm\,$0.03 &  14.86$\,\pm\,$0.01 &  14.15$\,\pm\,$0.01 &  13.74$\,\pm\,$0.01 &  13.34$\,\pm\,$0.01 \\
11 &  16.57$\,\pm\,$0.03 &  16.45$\,\pm\,$0.01 &  15.78$\,\pm\,$0.01 &  15.39$\,\pm\,$0.01 &  15.01$\,\pm\,$0.01 \\
12 &  15.60$\,\pm\,$0.03 &  15.75$\,\pm\,$0.01 &  15.38$\,\pm\,$0.01 &  15.09$\,\pm\,$0.01 &  14.80$\,\pm\,$0.01 \\
13 &  18.43$\,\pm\,$0.03 &  17.06$\,\pm\,$0.01 &  15.73$\,\pm\,$0.01 &  14.91$\,\pm\,$0.01 &  14.19$\,\pm\,$0.01 \\
14 &  16.15$\,\pm\,$0.03 &  16.03$\,\pm\,$0.01 &  15.35$\,\pm\,$0.01 &  14.97$\,\pm\,$0.01 &  14.56$\,\pm\,$0.01 \\
\noalign{\smallskip} \hline
\end{tabular}
\end{table*}
\begin{table*}
\centering
\renewcommand{\arraystretch}{1.1}
\caption{Optical photometry of \sniip\ from HCT.}
\label{tab:photlog}
\begin{tabular}{c c c c c c c c}
\hline \hline
       Date     &      JD       & Phase$^*$ &     $U$               &    $B$                &    $V$                &    $R$                &    $I$                \\
(yyyy-mm-dd)    & (2456700+)    &   (d)     &   (mag)               &   (mag)               &   (mag)               &   (mag)               &   (mag)               \\
 \hline \noalign{\smallskip}
 2014-07-15     &   154.31      &    +12.81 &       ---             &   16.05$\,\pm\,$0.03  &   15.94$\,\pm\,$0.03  &   15.72$\,\pm\,$0.03  &   15.52$\,\pm\,$0.03  \\
 2014-07-16     &   155.38      &    +13.88 &  15.54$\,\pm\,$0.06   &   16.04$\,\pm\,$0.02  &   15.96$\,\pm\,$0.02  &   15.73$\,\pm\,$0.02  &   15.55$\,\pm\,$0.03  \\
 2014-07-21     &   160.31      &    +18.81 &  15.93$\,\pm\,$0.05   &   16.24$\,\pm\,$0.02  &   16.02$\,\pm\,$0.02  &   15.75$\,\pm\,$0.03  &   15.58$\,\pm\,$0.02  \\
 2014-07-24     &   163.35      &    +21.85 &  16.26$\,\pm\,$0.05   &   16.42$\,\pm\,$0.02  &   16.09$\,\pm\,$0.02  &   15.78$\,\pm\,$0.02  &   15.61$\,\pm\,$0.02  \\
 2014-07-25     &   164.24      &    +22.74 &  16.35$\,\pm\,$0.05   &   16.45$\,\pm\,$0.02  &   16.09$\,\pm\,$0.02  &   15.77$\,\pm\,$0.02  &   15.58$\,\pm\,$0.02  \\
 2014-07-30     &   169.35      &    +27.85 &  16.85$\,\pm\,$0.04   &   16.75$\,\pm\,$0.02  &   16.18$\,\pm\,$0.02  &   15.83$\,\pm\,$0.02  &   15.62$\,\pm\,$0.02  \\
 2014-08-02     &   172.29      &    +30.79 &  17.12$\,\pm\,$0.04   &   16.94$\,\pm\,$0.02  &   16.27$\,\pm\,$0.02  &   15.90$\,\pm\,$0.02  &   15.70$\,\pm\,$0.02  \\
 2014-08-03     &   173.30      &    +31.80 &       ---             &   17.00$\,\pm\,$0.07  &   16.26$\,\pm\,$0.01  &   15.84$\,\pm\,$0.02  &   15.67$\,\pm\,$0.02  \\
 2014-08-08     &   178.21      &    +36.71 &  17.55$\,\pm\,$0.05   &   17.17$\,\pm\,$0.02  &   16.35$\,\pm\,$0.01  &   15.91$\,\pm\,$0.02  &   15.70$\,\pm\,$0.02  \\
 2014-08-09     &   179.12      &    +37.62 &       ---             &       ---             &       ---             &       ---             &   15.71$\,\pm\,$0.02  \\
 2014-08-12     &   182.21      &    +40.71 &  17.72$\,\pm\,$0.09   &  17.27$\,\pm\,$0.03   &   16.45$\,\pm\,$0.02  &   15.97$\,\pm\,$0.02  &   15.73$\,\pm\,$0.02  \\
 2014-08-13     &   183.42      &    +41.92 &       ---             &       ---             &   16.42$\,\pm\,$0.02  &   15.97$\,\pm\,$0.02  &   15.74$\,\pm\,$0.02  \\
 2014-08-27     &   197.19      &    +55.69 &       ---             &  17.63$\,\pm\,$0.02   &   16.61$\,\pm\,$0.02  &   16.10$\,\pm\,$0.02  &   15.87$\,\pm\,$0.02  \\
 2014-08-31     &   201.20      &    +59.70 &  18.61$\,\pm\,$0.04   &  17.70$\,\pm\,$0.02   &   16.62$\,\pm\,$0.01  &   16.11$\,\pm\,$0.02  &   15.86$\,\pm\,$0.02  \\
 2014-09-11     &   212.32      &    +70.82 &  18.84$\,\pm\,$0.08   &  17.92$\,\pm\,$0.02   &   16.74$\,\pm\,$0.01  &   16.21$\,\pm\,$0.02  &   15.94$\,\pm\,$0.02  \\
 2014-09-15     &   216.10      &    +74.60 &  19.23$\,\pm\,$0.05   &  18.02$\,\pm\,$0.02   &   16.79$\,\pm\,$0.02  &   16.29$\,\pm\,$0.04  &   15.98$\,\pm\,$0.02  \\
 2014-09-16     &   217.16      &    +75.66 &  19.30$\,\pm\,$0.05   &  18.05$\,\pm\,$0.02   &   16.82$\,\pm\,$0.01  &   16.26$\,\pm\,$0.02  &   15.99$\,\pm\,$0.02  \\
 2014-09-21     &   222.20      &    +80.70 &  19.44$\,\pm\,$0.06   &  18.17$\,\pm\,$0.02   &   16.89$\,\pm\,$0.01  &   16.31$\,\pm\,$0.02  &   16.05$\,\pm\,$0.02  \\
 2014-09-27     &   228.15      &    +86.65 &       ---             &  18.39$\,\pm\,$0.02   &   17.02$\,\pm\,$0.02  &   16.45$\,\pm\,$0.03  &   16.14$\,\pm\,$0.02  \\
 2014-10-08     &   239.06      &    +97.56 &  20.61$\,\pm\,$0.41   &  19.08$\,\pm\,$0.03   &   17.52$\,\pm\,$0.02  &   16.84$\,\pm\,$0.02  &   16.48$\,\pm\,$0.02  \\
 2014-10-09     &   240.23      &    +98.73 &       ---             &  19.19$\,\pm\,$0.04   &   17.66$\,\pm\,$0.02  &   16.93$\,\pm\,$0.02  &   16.56$\,\pm\,$0.02  \\
 2014-10-10     &   241.07      &    +99.57 &       ---             &  19.32$\,\pm\,$0.02   &   17.71$\,\pm\,$0.02  &   16.97$\,\pm\,$0.02  &   16.60$\,\pm\,$0.02  \\
 2014-10-20     &   251.26      &   +109.76 &       ---             &  20.26$\,\pm\,$0.05   &   18.66$\,\pm\,$0.07  &   17.83$\,\pm\,$0.03  &   17.40$\,\pm\,$0.03  \\
 2014-10-28     &   259.17      &   +117.67 &       ---             &  20.41$\,\pm\,$0.03   &   18.89$\,\pm\,$0.02  &   18.00$\,\pm\,$0.02  &   17.57$\,\pm\,$0.02  \\
 2014-11-04     &   266.13      &   +124.63 &       ---             &       ---             &   18.99$\,\pm\,$0.02  &   18.11$\,\pm\,$0.02  &   17.69$\,\pm\,$0.02  \\
 2014-11-17     &   279.18      &   +137.68 &       ---             &  20.57$\,\pm\,$0.03   &   19.13$\,\pm\,$0.02  &   18.25$\,\pm\,$0.02  &   17.84$\,\pm\,$0.03  \\
 2014-11-23     &   285.06      &   +143.56 &       ---             &  20.67$\,\pm\,$0.04   &   19.25$\,\pm\,$0.02  &   18.35$\,\pm\,$0.02  &   17.95$\,\pm\,$0.02  \\
 2014-12-01     &   293.09      &   +151.59 &       ---             &       ---             &   19.25$\,\pm\,$0.03  &   18.35$\,\pm\,$0.02  &       ---         \\
 2014-12-04     &   296.11      &   +154.61 &       ---             &       ---             &   19.33$\,\pm\,$0.03  &   18.43$\,\pm\,$0.02  &   18.00$\,\pm\,$0.02  \\
 2014-12-08     &   300.08      &   +158.58 &       ---             &       ---             &   19.26$\,\pm\,$0.02  &   18.43$\,\pm\,$0.02  &   18.07$\,\pm\,$0.02  \\
 2014-12-19     &   311.04      &   +169.54 &       ---             &       ---             &   19.20$\,\pm\,$0.02  &   18.46$\,\pm\,$0.03  &   18.07$\,\pm\,$0.02  \\
 2014-12-26     &   318.04      &   +176.54 &       ---             &       ---             &   19.52$\,\pm\,$0.02  &   18.59$\,\pm\,$0.02  &   18.26$\,\pm\,$0.03  \\
 2015-01-03     &   326.04      &   +184.54 &       ---             &       ---             &   19.69$\,\pm\,$0.03  &   18.76$\,\pm\,$0.03  &   18.40$\,\pm\,$0.03  \\
 2015-01-25     &   348.07      &   +206.57 &       ---             &       ---             &   19.80$\,\pm\,$0.04  &   18.78$\,\pm\,$0.03  &   18.58$\,\pm\,$0.04  \\
\noalign{\smallskip} \hline
\multicolumn{3}{l}{$^*$\footnotesize{Time since explosion epoch (JD 2456841.50)}.}
\end{tabular}
\end{table*}
\begin{table*}
\caption{Distances to the host galaxy UGC 11860.}
\renewcommand{\arraystretch}{1.6}
\label{tab:dist_host}
\begin{tabular}{l c c c}
  \hline \hline
  Distance Method					                & Distance 	            & Distance Modulus      	    &   Reference 	\\
                                                    & (Mpc)                 &   (mag)                       &               \\
  \noalign{\smallskip} \hline
  Galactocentric 					                & 45.7$\,\pm\,$3.2		& 33.30$\,\pm\,$0.15		    & 	2	        \\
  Local Group 					                    & 46.8$\,\pm\,$3.3		& 33.35$\,\pm\,$0.15		    & 	2	        \\
  CMB Dipole Model 					                & 38.2$\,\pm\,$2.7		& 32.91$\,\pm\,$0.15	    	&	2	        \\
  Virgo Infall only					                & 45.7$\,\pm\,$3.2		& 33.30$\,\pm\,$0.15		    &	2	        \\
  Virgo + Great Attractor Infall only		        & 44.9$\,\pm\,$3.1 		& 33.26$\,\pm\,$0.15		    &	2	        \\
  Virgo + Great Attractor + Shapley Supercluster	& 44.9$\,\pm\,$3.1  	& 33.26$\,\pm\,$0.15		    &	2	        \\
  Standard Candle Method				            & 47.2$\,\pm\,$3.6  	& 33.37$\,\pm\,$0.17		    &	1	        \\ \noalign{\smallskip} \hline
  Mean Distance					                    & 44.8$\,\pm\,$3.0  	& 33.25$\,\pm\,$0.15		    &		        \\ \noalign{\smallskip} \hline
 \end{tabular}
\newline
(1) This paper;
(2) NASA/IPAC Extragalactic Database (NED).
\end{table*}

\section{Figures}

\begin{figure*}
\centering
\resizebox{\hsize}{!}{\includegraphics{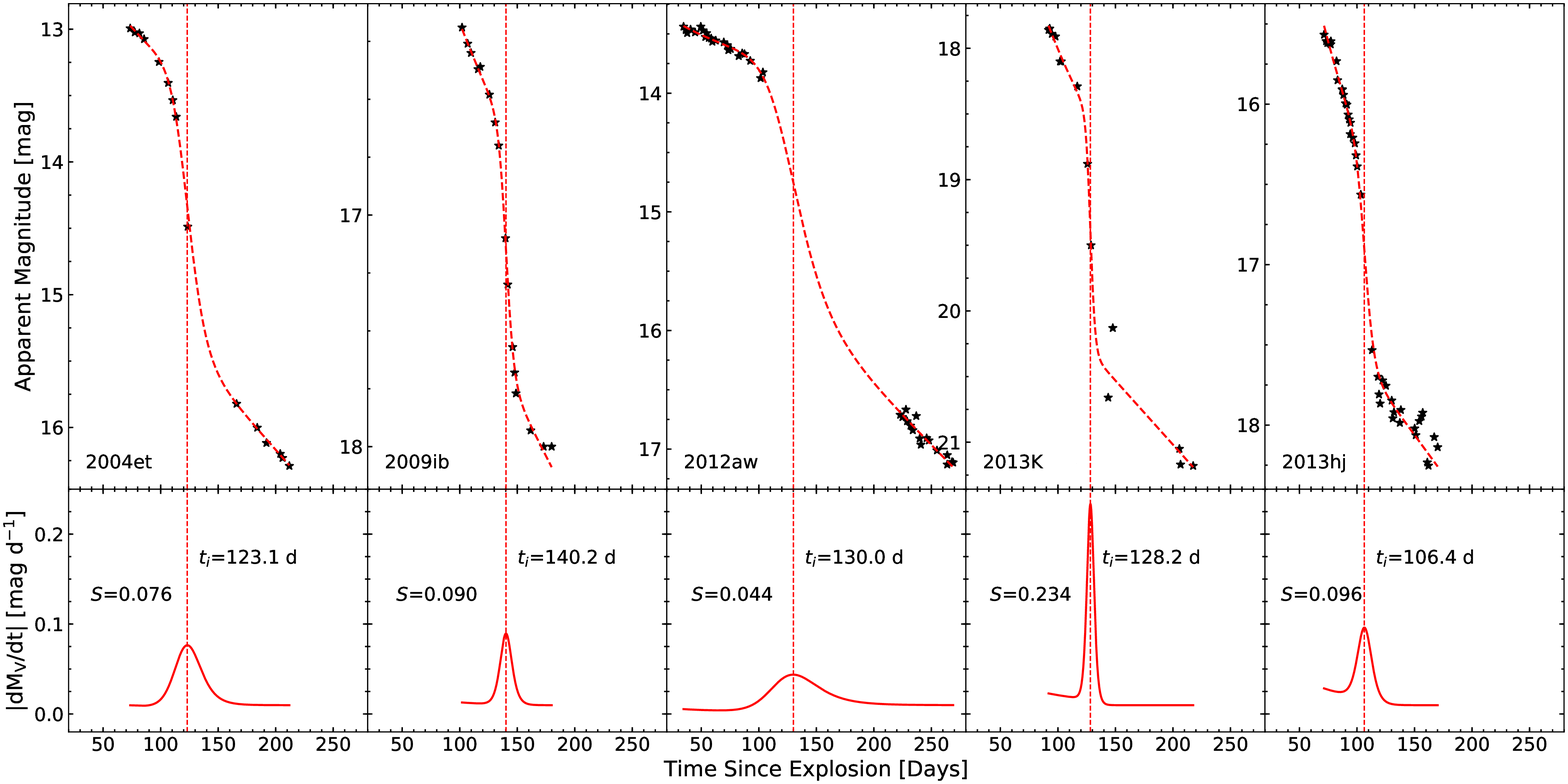}}
\caption{Determination of Steepness parameter and point of inflection for type II SNe 1992ba, 2004et, 2009ib, 2012aw and 2013K. The plot description is same as in Fig.~\ref{fig:steepness}.}
\label{fig:miscsteepness1}
\end{figure*}

\begin{figure*}
\centering
\resizebox{\hsize}{!}{\includegraphics{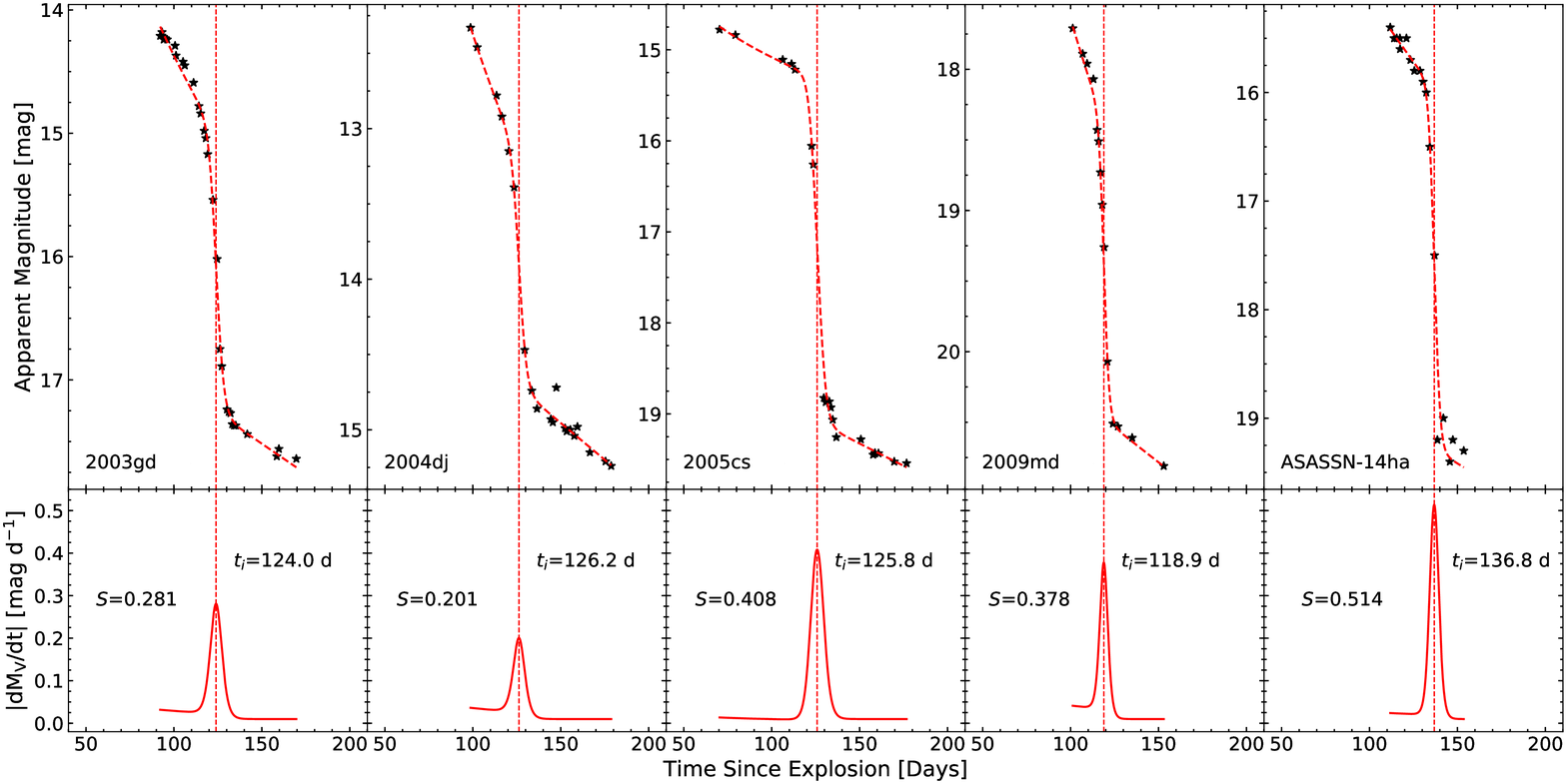}}
\caption{Determination of Steepness parameter and point of inflection for type II SNe 2003gd, 2004dj, 2005cs, 2009md and ASASSN-14ha. The plot description is same as in Fig.~\ref{fig:steepness}.}
\label{fig:miscsteepness2}
\end{figure*}

\begin{figure*}
\centering
\resizebox{\hsize}{!}{\includegraphics{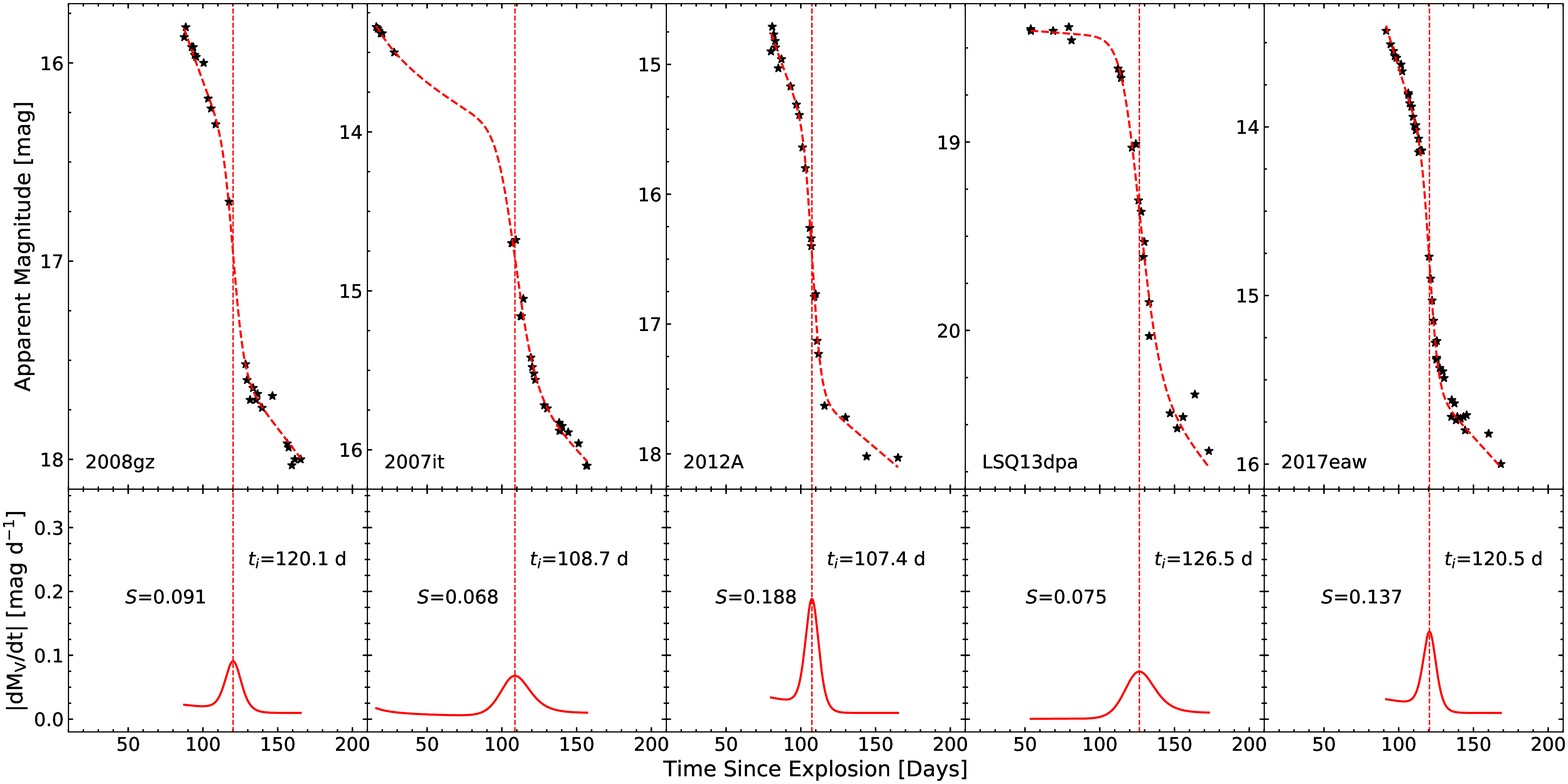}}
\caption{Determination of Steepness parameter and point of inflection for type II SNe 2008gz, 2007it, 2012A, LSQ13dpa and 2017eaw. The plot description is same as in Fig.~\ref{fig:steepness}.}
\label{fig:miscsteepness3}
\end{figure*}

\begin{figure*}
\centering
\resizebox{\hsize}{!}{\includegraphics{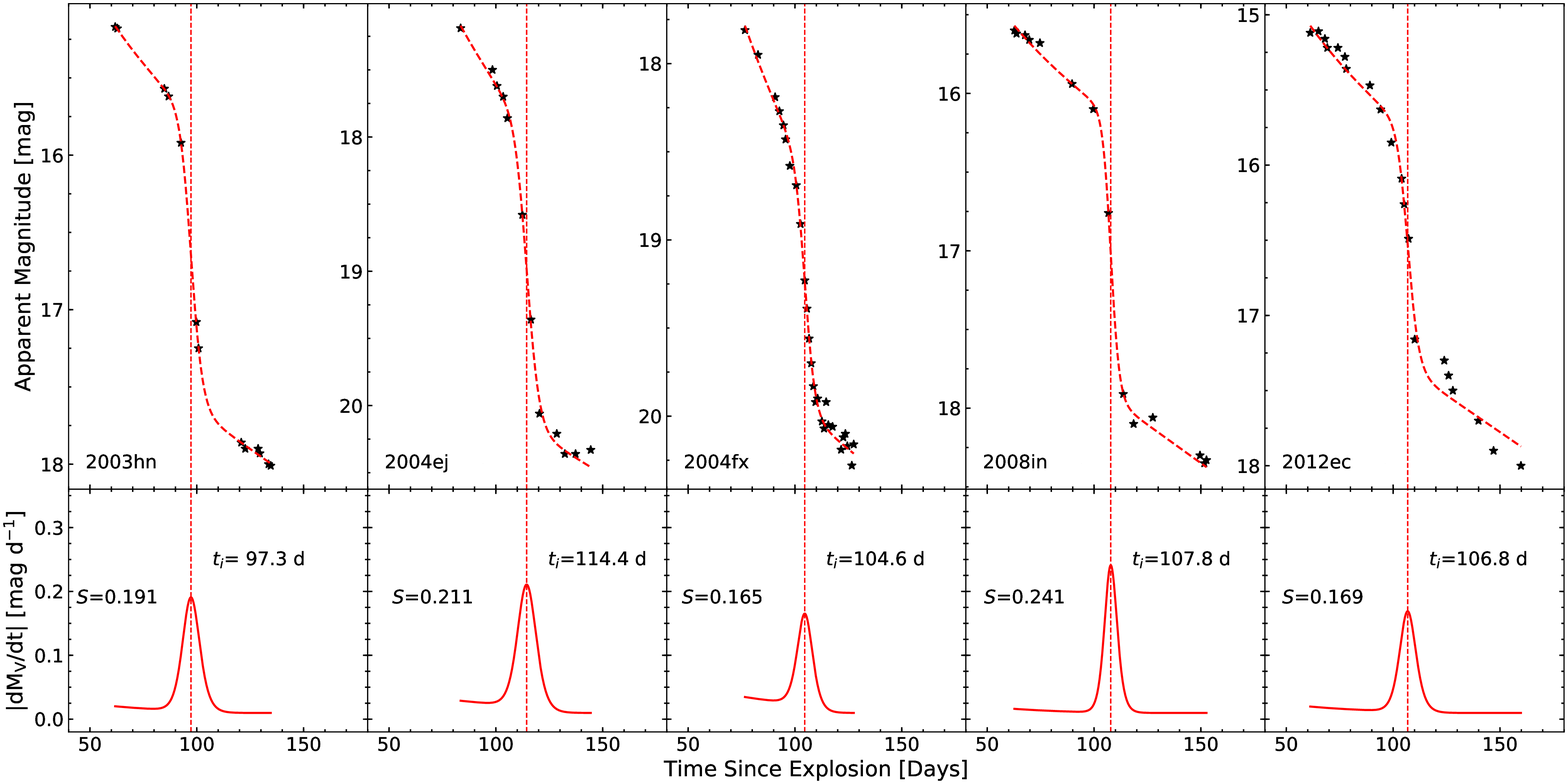}}
\caption{Determination of Steepness parameter and point of inflection for type II SNe 2003hn, 2004ej, 2004fx, 2008in and 2012ec. The plot description is same as in Fig.~\ref{fig:steepness}.}
\label{fig:miscsteepness4}
\end{figure*}

\begin{figure*}
\centering
\resizebox{\hsize}{!}{\includegraphics{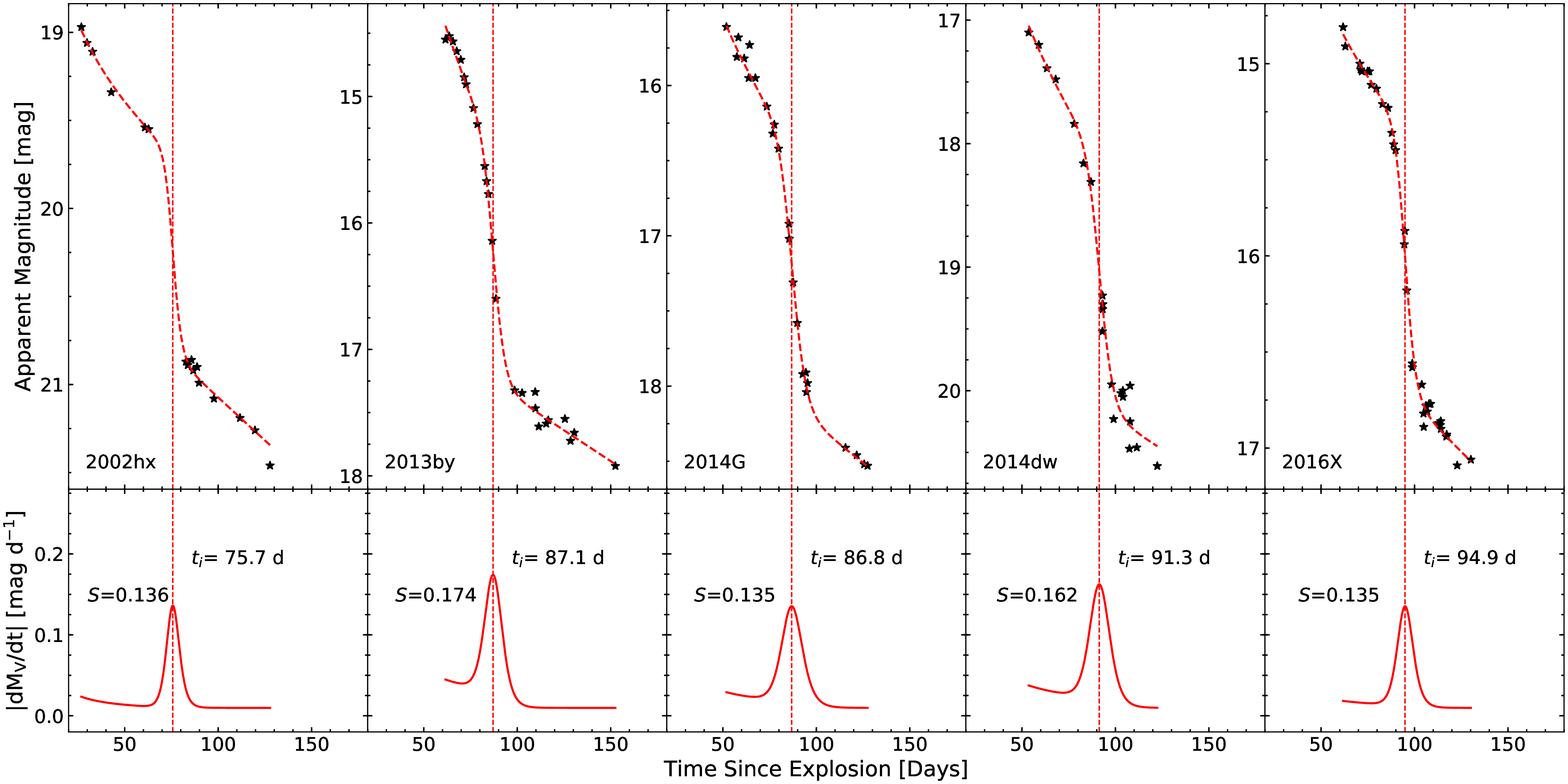}}
\caption{Determination of Steepness parameter and point of inflection for type II SNe 2002hx, 2013by, 2014G, 2014dw and 2016X. The plot description is same as in Fig.~\ref{fig:steepness}.}
\label{fig:miscsteepness5}
\end{figure*}

\end{document}